\documentclass[fleqn,usenatbib]{mnras}

\usepackage{newtxtext,newtxmath}

\usepackage[T1]{fontenc}

\DeclareRobustCommand{\VAN}[3]{#2}
\let\VANthebibliography\thebibliography
\def\thebibliography{\DeclareRobustCommand{\VAN}[3]{##3}\VANthebibliography}

\usepackage{graphicx}	
\usepackage{amsmath}	
\usepackage{longtable}  
\usepackage{lineno}

\newcommand*{\var}{\textcolor{black}{$\sim$}}
\newcommand*{\none}{\textcolor{black}{$\times$}}

\title[Pulse profile modeling of thermonuclear burst oscillations I]{Pulse Profile Modeling of Thermonuclear Burst Oscillations I: The Effect of Neglecting Variability}

\author[Kini~et~al.]{Yves~Kini$^{1}$
\thanks{E-mail: \href{mailto:y.kini@uva.nl}{y.kini@uva.nl}},
Tuomo~Salmi$^{1}$, 
Anna~L.~Watts$^{1}$,
Serena~Vinciguerra$^{1}$,
Devarshi~Choudhury$^{1}$,\newauthor 
Siem~Fenne$^{1}$,  
Slavko~Bogdanov$^{2}$, 
Zach~Meisel$^{3}$ and
Valery~Suleimanov$^{4}$ 
\\
$^{1}$Anton Pannekoek Institute for Astronomy, University of Amsterdam, Science Park 904, 1090GE Amsterdam, the Netherlands\\
$^{2}$Columbia Astrophysics Laboratory, Columbia University, 550 West 120th Street, New York, NY 10027, USA\\
$^{3}$Institute of Nuclear \& Particle Physics, Department of Physics \& Astronomy, Ohio University, Athens, Ohio 45701, USA\\
$^{4}$Institut f\"ur Astronomie und Astrophysik, Kepler Center for Astro and Particle Physics, Universit\"at T\"ubingen, Sand 1, D-72076 T\"ubingen, Germany
}

\date{Accepted XXX. Received YYY; in original form ZZZ}

\pubyear{20XX}

\begin{document}
\label{firstpage}
\pagerange{\pageref{firstpage}--\pageref{lastpage}}
\maketitle

\begin{abstract}

We study the effects of the time-variable properties of thermonuclear X-ray bursts on modeling their millisecond-period burst oscillations. We apply the pulse profile modeling technique that is being used in the analysis of rotation-powered millisecond pulsars by the Neutron Star Interior Composition Explorer (NICER) to infer masses, radii, and geometric parameters of neutron stars. 
By simulating and analyzing a large set of models, we show that overlooking burst time-scale variability in temperatures and sizes of the hot emitting regions can result in substantial bias in the inferred mass and radius. 
To adequately infer neutron star properties, it is essential to develop a model for the time variable properties or invest a substantial amount of computational time in segmenting the data into non-varying pieces. We discuss prospects for constraints from proposed future X-ray telescopes. 

\end{abstract}

\begin{keywords}
dense matter --- equation of state --- pulsars: general --- pulsars: individual (XTE~J1814$-$338) --- stars: neutron --- X-rays: stars
\end{keywords}



\section{Introduction}\label{sec:intro}

Neutron stars (NSs) are the densest directly observable objects in the universe, which  makes their cores the ideal environment to probe the properties of matter at high density and low temperature. Probing the matter properties in the core can be achieved by measuring their mass and radius, which depend on the dense matter Equation of State (EoS) \citep[see e.g.][]{ Lattimer:2012, Oertel:2016, Baym:2017, Tolos:2020, Yang:2019, Hebeler:2020}.

For X-ray pulsars, mass and radius measurement is  possible through analysis of X-ray pulsations that emanate from their surface.  These coherent X-ray pulsations are produced by hot spots as they rotate in and out of view as the star spins. The intense gravitational field experienced by photons as they leave the stellar surface, and the fast rotation of the star, shape the characteristics of the observed  pulsations. Therefore, by using pulse profile modeling, a technique that combines special and general relativity, it is possible to infer the stellar mass and radius in addition to other key parameters \citep[see e.g][and references therein]{Watts:2019, Bogdanov:2019}.

To successfully infer the stellar properties, arriving photons from an actual source are first binned by energy and rotational phase, according to an ephemeris, to produce an observed pulse profile. Then a simulated pulse profile for a given set of input parameters is generated using mathematical models that take into account the relevant physical phenomena and parameters, such as the pulsar's rotation, magnetic field, and the emission and propagation of electromagnetic radiation. Next, the simulated pulse profile is compared to observations of the pulsar's actual pulse profile, using statistical and other analysis techniques to assess the degree of agreement between the two. Finally, the model and input parameters are adjusted if necessary, based on the comparison between the simulated and observed pulse profiles, in order to improve the accuracy of the model. This process is then repeated until the simulated pulse profile closely matches the observed pulse profile and gives a good picture of the physical processes that cause pulsar emission. 

Recently, the pulse profile modeling technique has been used to place stringent parameter constraints on some of the Neutron Star Interior Composition Explorer \citep[NICER;][]{NICER} primary targets; PSR J0030$+$0451 \citep{Riley:2019, Miller:2019} and PSR J0740$+$6620 \citep{Riley:2021, Miller:2021, Salmi:2022}. NICER has constrained not only the masses and radii of these Rotation-Powered Millisecond Pulsars (RMPs), but also the hot spot properties and by extension the magnetic field geometry \citep{Bilous:2019, Chen:2020, Kalapotharakos:2021}. However pulse profile modeling can also be applied to Accretion-Powered Millisecond Pulsars (AMPs) and Thermonuclear Burst Oscillation (TBO) sources \citep[see for example][]{Watts:2016}. The latter are the primary focus of this paper.

First observed in 1976 \citep{Grindlay:1976, Belian:1976}, Type I X-ray bursts are thermonuclear runaways caused by unstable burning of the accreted plasma onto the NS surface from its companion. This unstable burning occurs under specific conditions of accretion rate and chemical composition of the accreted material \citep{Bildsten:1998, Galloway:2017, Galloway:2020}. The typical duration of the most common bursts varies from a few to several hundred seconds. Their light curves exhibit a fast rise in luminosity followed by an exponential decay\footnote{Superbursts, which are believed to be caused by carbon combustion, can last for several hours and generate energy of up to $10^{42}$ ergs \citep[see e.g.][for review]{Zand:2017}.}. Their recurrence time ranges from a few minutes to days, and the energy released during a burst is typically $10^{39}$-$10^{40}$ ergs. Two decades after the first burst was discovered, coherent pulsations, commonly known as thermonuclear burst oscillations or simply burst oscillations, were found in some of the burst light curves of 4U $1728-34$ \citep{Strohmayer:1996}. These coherent pulsations have now been firmly confirmed in nearly 20$\%$ of thermonuclear bursters \citep{Bilous:2018fxm}. 

All known TBO sources have millisecond periods, with the exception of IGR J17480-2446 \citep[$\sim$11 Hz;][]{Strohmayer:2010, Altamirano:2010}. Their spin rates are typically between $\sim$ 250 Hz and $\sim$ 600 Hz. Rapid rotation enhances a variety of effects, namely the Doppler boost of photon energy, the angular distribution aberration, and the time delay between photons emitted at different phases. These effects introduce asymmetry and harmonic content in the observed pulse profile, which combined with the compactness information gained from the pulse amplitude, can be utilized to independently determine the mass and the radius through pulse profile modeling \citep[see e.g.][ for discussion]{Braje:2000, Poutanen:2006, Lo:2013ava,Psaltis:2013fha}.  A benefit of using TBO sources is that the inferred mass and radius can be independently cross-checked using other methods, for example, through pulse profile modeling of the accretion pulsations of AMPs \citep[e.g.][]{Salmi:2018} or by modeling the burst spectra \citep[e.g.][]{Suleimanov:2011,Ozel:2013, Steiner:2013, Guillot:2014, Nattila:2016, Steiner:2018}.

Still, attempting to apply pulse profile modeling to burst oscillation sources comes with multiple challenges. First, where exactly these pulsations originate from is still up for debate (see Sec \ref{sec:origins}). Models for surface patterns are required but the uncertainties associated with them are likely to influence the inferred source properties \citep{Watts:2012, Bhattacharyya:2022}. The second challenge stems from the burst's inherent variability, as well as the variability associated with the oscillations. Bursts are by definition variable, as temperature evolution due to thermonuclear burning and then cooling drives the fast increase and then slower decrease in X-ray flux. The presence of oscillations may however lead to additional sources of variability in the physical emission model (e.g. relating to hot spot motion or size). 
If not correctly addressed, these variabilities could lead to biased inferred parameters. The third difficulty associated with bursters is the low root mean square (rms) Fractional Amplitude (FA). This increases the already wide number of parameter degeneracies intrinsic to pulse profile modeling \citep[see e.g][]{Lo:2013ava, Psaltis:2013fha}. The final obstacle comes from the bursting NS atmosphere. X-ray photons from the stellar surface of the star are reprocessed in the atmosphere and this must be accounted for. The current state-of-the-art atmosphere models \citep{Valery2012, Valery2018} for TBO sources make assumptions about the composition of the atmosphere, among other things. Since the actual composition of the atmosphere is not known, these assumptions leave room for modeling uncertainty.
Although applying pulse profile modeling to TBO sources is rendered challenging by the aforementioned factors, it may not only shed light on the fundamental interactions taking place in the NS cores, but also provide a unique opportunity to explore the climate and weather of bursters and  study the physics of nuclear burning.

The purpose of this paper is to determine whether time variation in temperature and/or emitting region radius during the burst can be neglected, which would reduce computational costs in analysis. Therefore, we make  
use of simulations to assess how well parameters can be recovered, adapting the X-ray Pulse Simulation and Inference \citep[X-PSI;][]{Riley2023} pipeline developed for NICER analysis to model burst oscillations. Our simulations must account for the lack of knowledge of the spot pattern, which introduces a lot of uncertainty. To mitigate this, we use cutting-edge nuclear burning models and tailor our analysis to the best source in this class, XTE~J1814$-$338 \citep[hereafter J1814;][]{Strohmayer:2003, Bhattacharyya:2004pp}.

Our analysis proceeds in two stages. Firstly, using pulse profile modeling and assuming reasonable possible surface patterns, we construct two sets of synthetic data; with and without time variation in temperature and/or emission region radius. We then perform inference runs on the various sets of synthetic data, ignoring the temporal variation for both sets. Thus, we structure this paper as follows. First, we describe in Section \ref{sec:method}, our modeling approach, assumptions, and synthetic data generation method. Then, we present the results of our analysis in Section  \ref{sec:result}. Finally we discuss the implications of using pulse profile modelling for TBO sources in Section  \ref{sec:discussion}.

\section{METHOD }\label{sec:method} 


Photons contained in X-ray pulsations captured by a telescope carry fundamental information about their emitting sources, interactions in the various media they traversed, and the method by which they were recorded \citep{Pechenick:1983,Chen:1989,Page:1995,Miller:1998,Braje:2000, Weinberg:2001,Beloborodov:2002, Poutanen:2006, Cadeau:2007,Morsink:2007, Baubock:2012,Lo:2013ava, Psaltis:2013fha, Miller:2015,Stevens:2016, Nattila:2018, Bogdanov:2019}. Retrieving this information through pulse profile modelling requires: a surface pattern model, an atmosphere model, a model for the interstellar medium (ISM) extinction and the instrument response. In addition, a mathematical model of the NS's external gravitational field and relativistic effects are required to model the propagation of the emitted photons. We employ the oblate-star Schwarzschild space-time plus Doppler approximation \citep{Cadeau:2007, Morsink:2007}, which is adequate for the spin rates of known TBO sources, and use it for relativistic ray tracing. For the instrument response, we make use of the Rossi X-Ray Timing Explorer (RXTE), Proportional Counter Array \citep[PCA;][]{Jahoda:1996} instrument response file\footnote{We used the response matrix of burst 10 of J1814 (ObsID: 80418-01-02-00, MJD: 52803.7406). All the five proportional counter units were included when generating the response file.}. All the surface patterns, relativistic ray tracing, and accounting for the ISM, the NS atmospheric interactions  and the instrument response are done using our publicly available simulation and inference package, X-PSI\footnote{\url{https://github.com/xpsi-group/xpsi.git}} \citep{Riley2023}, version v0.7.9\footnote{During sampling, we slightly modified the $eval\_marginal\_likelihood$ function to bypass the check that determines how many sigmas below the signal from the star the data can be (at any channel) before skipping the precise likelihood computation. This requirement was deemed too restrictive for bursting sources. Some X-PSI version 1.0.0 \citep{xpsi_v122_zenodo} features were also used for postprocessing.}. X-PSI uses the Bayesian inference algorithm  \texttt{MultiNest} \citep{MultiNest_2008,MultiNest_2009, MultiNest_2019} to explore the large parameter space and infer the source properties.

In this section, we discuss all of the modeling components, from photon emission to the instrument. We begin by describing possible pulsation origins, our atmosphere and ISM assumptions. Since our pulse profiles include a background component, we explain how this was incorporated into the modeling. Finally, we discuss the  prior assumptions required for the \texttt{MultiNest} Bayesian framework.

\subsection{Oscillation origins and phenomenological models}\label{sec:origins}

For RMPs, pulsations are caused by hot spots induced by return currents at the magnetic poles \citep{Harding:2001,Harding:2002}. Despite the well-known origin of the pulsations, the resulting surface pattern is uncertain \citep{Gralla:2017,Philippov:2018}, resulting in temperature distributions being parameterized with simple models in NICER's RMPs analyses \citep{Riley:2019, Miller:2019, Riley:2021, Miller:2021, Salmi:2022}.

For TBO sources however, the origin of the pulses is still unclear \citep{Watts:2012, Bhattacharyya:2022}.  In the presence of a high magnetic field and/or rapid rotation, fuel and/or flame confinement may cause localized burning \citep{Cavecchi:2011}. This confined burning would result in a hotter region, which would explain the observed oscillations during the rise and the decay of the burst. Yet, not all sources have a significant magnetic field or sufficiently rapid rotation to confine the flame for long. For those sources, the flames would spread from their ignition points, inducing oscillations during the rise \citep{Strohmayer:1997}. In this scenario, the amplitude of the pulsation and their presence in the tail of the burst are more difficult to explain. The explanation may potentially come from global modes, or more precisely, buoyant r-modes \citep{Heyl:2004, Lee:2004, Piro:2005,Chambers:2019, Chambers:2020}, a cooling wake \citep{Mahmoodifar:2016}, convective patterns \citep{Garcia:2018}, or the dynamics of vortices \citep{Spitkovsky:2002,Cavecchi:2019}.

Given the uncertainties related to oscillation origins and hence the ``true'' form of the surface pattern, we employ a phenomenological approach, similar to NICER's RMP analysis, and tailor our analysis to J1814. J1814 is one of the best sources for pulse profile modeling due to its very stable  pulsations (very little frequency drift), relatively high FA, and harmonic content in the pulsations during and between bursts \citep{Strohmayer:2003, Bhattacharyya:2004pp,Watts:2006}. The stable  pulsations help to minimize the complexity involved with the TBO source. Furthermore, high FA and harmonic content are expected to help overcome the degeneracy between some parameters, particularly the mass and radius \citep{Psaltis:2013fha}.

We selected phenomenological models that allow us to reproduce some basic key properties of the light curve and oscillations of J1814, assuming different surface temperature pattern properties.
We developed eight models and following is an overview  of each model, grouped by surface pattern and summarized in Table \ref{tab:models}. To conform to the naming patterns of earlier X-PSI models \citep{Riley:2019}, we will prefix each model name with the symbol ST-, which stands for Single Temperature. This is mainly  because we assume a single hot spot, emitting at a uniform temperature,  and do not yet consider the possibility of more than one emitting hot region. While our choice of models is necessarily restricted,  they nevertheless allow us to simulate variability of hot spot properties and test the effects on parameter recovery.

ST-$\tilde{H}_{T}$ \& ST-$\bar{S}\tilde{H}_{T}$: Both models assume a hot spot on the surface of a star with a set angular radius and a variable temperature over time. Such a scenario would be consistent with confined burning, with fuel directed to a small region of the star by a strong magnetic field, or because the combined effects of a magnetic and Coriolis force would prevent the flame from spreading. With ST-$\tilde{H}_{T}$, we presume that the star's non-burning region is too cool to emit X-rays in the bands covered by the instrument's effective area. Model ST-$\bar{S}\tilde{H}_{T}$, however, allows the rest of the star to emit with a uniform and constant temperature.

ST-$\tilde{H}_{T,R}$ \& ST-$\bar{S}\tilde{H}_{T,R}$: Similar to previous models, these models assume a temperature-varying hot spot, but this time, we also allow the burning region angular radius to evolve with time. 
These models would be consistent with a scenario in which the flames propagate outward from the ignition point but ultimately stall due to the interaction of the magnetic and/or Coriolis forces. Similar to earlier models, model ST-$\tilde{H}_{T,R}$ prevents the non-burning region of the star from emitting, whereas ST-$\bar{S}\tilde{H}_{T,R}$ allows this to occur. 

 ST-$\tilde{S}_{T}\bar{H}$ \& ST-$\tilde{S}_{T}\tilde{H}_{R}$:  These two models allow the bulk of the star's surface temperature to vary over time, with the exception of an isolated hot spot with a constant temperature. This would correspond to a scenario where the flame spreads and engulfs the entire NS surface. As for the origin of the hot spot responsible for the pulsations, we speculate that it may have existed before the explosion (e.g. accretion spot), or that it may have been produced by some unknown mechanism on the star surface, in the atmosphere, or in the accretion funnel. In the ST-$\tilde{S}_{T}\tilde{H}_{R}$ model, the angular radius of this hotter region can extend over time, whereas in model  ST-$\tilde{S}_{T}\bar{H}$, this is not permitted. 

ST-$\tilde{S}_{T}\tilde{H}_{T}$ \& ST-$\tilde{S}_{T}\tilde{H}_{T,R}$: These two models resemble the two preceding them in term of burning occurring on the surface and origin of the oscillations.  Instead of a hot spot with a constant temperature, we allow this region to have a time-dependent temperature. Again, we speculate that variations in temperature over time occurring on the stellar surface may also result in variations in the temperature of the hot spot.

We utilize these eight distinct models to construct the \textit{variability data set} (see Section \ref{subsec:synthetic_data}), but during inference, we overlook any temporal variation in temperature and/or angular radius. Inference is based on the assumption that the temperature and angular radius remain constant throughout the burst.

    \begin{table}
    
    \begin{center}
    \caption{Table summarizing the models.  $T_{\mathrm{spot}}$, $\zeta_{\mathrm{spot }}$ and $T_{\mathrm{star}}$  are respectively the hot spot temperature, hot spot angular radius and star temperature. $\sim$, $-$ and $\times$ mean respectively that the corresponding  parameter varies over time, is constant and is non-existent or not considered for this model.}
    \begin{tabular}{p{2.cm} p{2.cm} p{2.cm} p{1.cm}}

        \hline \hline
            Models                            & $T_{\mathrm{spot}}$                            &$\zeta_{\mathrm{spot }}$                        & $T_{\mathrm{star}}$                             \\ \hline
             ST-$\tilde{H}_{T}$               & $\sim$                                &$-$                                    & $\times$                                     \\ 
             ST-$\tilde{H}_{T,R}$             & $\sim$                                &$\sim$                                 & $\times$                                     \\
             ST-$\bar{S}\tilde{H}_{T}$        & $\sim$                                &$-$                                    & $-$                                          \\
             ST-$\bar{S}\tilde{H}_{T,R}$      & $\sim$                                &$\sim$                                 & $-$                                          \\
             ST-$\tilde{S}_{T}\tilde{H}_{T}$  & $\sim$                                &$-$                                    &$\sim$                                        \\
             ST-$\tilde{S}_{T}\tilde{H}_{T,R}$&$\sim$                                 &$\sim$                                 &$\sim$                                         \\
             ST-$\tilde{S}_{T}\bar{H}$        & $-$                                   & $-$                                   &$\sim$                                         \\
             ST-$\tilde{S}_{T}\tilde{H}_{R}$  & $-$                                   &$\sim$                                 &$\sim$                                         \\
              
            \hline
         \end{tabular}
        
         \label{tab:models}
         \end{center}
         \end{table}

        \subsection {Atmosphere model}\label{subsubsec:atmosphere}
        
         As mentioned in Section \ref{sec:intro}, NSs have a thin  atmosphere layer at their surface, which needs to be understood to infer the stars' properties.  
        As photons travel through this surface layer, they are absorbed, re-emitted, and up-scattered to different energies and directions.  This causes the emergent spectra of thermonuclear bursts (TBs) to differ from isotropic black-body emission \citep[see e.g.][]{London:1984, London:1986, Lapidus:1986, Ebisuzaki:1987, Madej:1991,Madej:2004, Valery2011,Valery2012,Worpel:2013fka,Valery2017,Valery2020}. 
        In this work, we make use of the atmosphere model developed for TBs in \citet{Valery2012}. 
        This model solves the radiation transfer, hydrostatic equilibrium, and energy balance equations using the exact relativistic Klein-Nishina cross-section and redistribution function for Compton scattering. 
        The spectra predicted by this model can be approximated closely by a diluted black-body; $F_E\approx w\pi B_E(f_{\mathrm{c}}\times T_{\mathrm{eff}})$ with dilution $w$ and colour-correction $f_{\mathrm{c}}$ factors depending entirely on the surface gravity, luminosity and chemical composition. The angular distribution of the emergent radiation is found to be well approximated by the Chandrasekhar-Sobolev beaming function for an electron scattering dominated atmosphere \citep{Cha47,sob49,Cha60,Sob63}: $I(\mu) \sim 1 + 2.06 \mu $, where $\mu$ is the cosine of the emission angle \citep{Valery2012,Valery2020}. 
        Thus, for simplicity, we fitted the atmosphere model spectra with a diluted blackbody using the same procedure as in \citet{Valery2012}, and calculated the total emitted intensity as $I_E(\mu)=B_{E}(f_{\mathrm{c}} T_{\mathrm{eff}})(0.421+0.868\mu)$, similarly to Equation (9) in \citet{Valery2020}.

        For a computationally efficient pulse profile calculation, a grid of atmosphere models were precomputed into a lookup table. 
        The format of this table is similar to those used in NICER-analysis and explained in \citet{Bogdanov:2021}, except for the fact that our surface gravities $\log (g/1\mathrm{cm\,s^{-2}})$ range from 13.7 to 14.9 with $\Delta \log (g/1\mathrm{cm\,s^{-2}}) = 0.1$ spacing and effective temperatures range from 6.7 to 7.6 (in $\log(T_{\mathrm{eff}}/1\mathrm{K})$) with $\Delta \log(T_{\mathrm{eff}}/1\mathrm{K}) = 0.02$ spacing \footnote{However, we note that the original atmosphere grid was defined in luminosities relative to the Eddington luminosity $l=L/L_{\mathrm{Edd}}$ with the following values: 0.001, 0.003, 0.01, 0.03, 0.05, 0.07, 0.1, 0.15, 0.2, 0.3, 0.4, 0.5, 0.55, 0.6, 0.65, 0.7, 0.75, 0.8, 0.85, 0.9, 0.95, 0.98, 1.0, 1.02, 1.04, 1.06, 1.08 ,1.1, 1.12.; and in surface gravities $\log (g/1\mathrm{cm\,s^{-2}})$ ranging from 13.7 to 14.9 with $\Delta \log (g/1\mathrm{cm\,s^{-2}}) = 0.15$ spacing. A new grid, defined as explained in the main text, was produced by interpolating to the corresponding temperature and surface gravity values.}.
        We also created a grid in photon energies and emission angles similar to that in \citet[][]{Bogdanov:2021} (i.e., from $-1.3$ to $2.0$ in $\log(E/kT_{\mathrm{eff}})$ and from $0$ to $1$ in $\mu$)
        , in order to easily apply the same interpolation methods as in the previous NICER-analysis when estimating the intensity for a given temperature, surface gravity, photon energy, and emission angle. 
        Throughout this analysis, we have applied atmosphere models with solar abundance composition ($X=0.737,\ Z=0.0134$) for synthetic data generation and statistical inference.

        \subsection {Interstellar absorption}
        As X-ray photons pass through the  interstellar medium (ISM), they are scattered and/or absorbed. This results in the initial spectrum being altered. The observed spectrum can be expressed as: $I_\mathrm{observed}=I_\mathrm{source}e^{-\sigma_\mathrm{ISM}(E)N_{H}}$, with $\sigma_\mathrm{ISM}(E)$ the  energy dependent ISM photoionization cross section normalised to the total hydrogen column density and $N_{H}$  the hydrogen column density measured in atoms cm$^{-2}$. We precomputed lookup tables of attenuation factors; $f_\mathrm{a}(E,N_{H})=e^{-\sigma_\mathrm{ISM}(E)N_{H}}$ as a function of the energy (ranging from 0.1 to 30 keV) using the X-ray absorption model \texttt{TBabs} \citep{Wilms:2000} for different hydrogen column densities. We find that, given a reference hydrogen column density $N_{H}^{\mathrm{ref}}$, $f_{\mathrm{a}}(E,N_{H}) \approx f_{\mathrm{a}}(E,N_{H}^{\mathrm{ref}})^{N_{H}/N_{H}^{\mathrm{ref}}}$ and verified that this  approximate relation provides sufficient accuracy for our analysis. Thus, for improving the computing efficiency particularly during sampling, we extract the attenuation factors using the scaling relation and use the attenuation factor corresponding to  $4\times 10^{20}$ atoms cm$^2$  as a reference hydrogen column density.

        \subsection {Background}
        
        To effectively estimate parameters using pulse profile modeling, a thorough understanding of the background distributions in the data is crucial \citep[see][]{Salmi:2022}. For burst oscillation sources, there are two key components that are expected to produce the background photons \citep{Watts:2019lbs}. The first, unpulsed, component originates from different X-ray sources in the field of view, the instrument electronics \citep[see e.g.][for a review]{Campana:2022}, and most significantly the observed source's close environment (e.g. accretion disk, accretion column, etc.). The second component, this time pulsed, may result from X-ray emission caused by accretion plasma striking the stellar surface (giving rise to accretion-powered pulsations). However, only eight of the nineteen known TBO sources have exhibited accretion-powered pulsations (persistent accretion-powered pulsations for six sources and intermittent accretion-powered pulsations for the remaining two). With the typical accretion hot spot temperature being $\sim 1$ keV \citep{Poutanen:2003, Salmi:2018} and the temperature at the  peak of the burst being $\sim 3$ keV \citep{Galloway:2020}, it is expected that the contribution of the accretion-powered pulsations will be low, especially at the peak of the burst \citep[see also][]{Watts:2005}.  Extra contributions to the pulsed background may originate from shocks in the accretion funnel as well as the pulsed X-ray signal from the stellar surface reflecting from the accretion disk. Given our current understanding of the origin of the burst oscillation, the pulsed component may or may not be categorized as background, depending on the phenomenological model considered.

         For the sake of simplicity, while generating the synthetic data, we introduced only an unpulsed  background with a black-body and power-law component. We made use of the  black-body and and power-law parameters values obtained by \citet{Krauss:2005sj} through spectral analysis of J1814. Hence, the black-body temperature and radius was fixed to 0.95 keV  and 1.6 km respectively, while the power-law index was set to 1.41 with an amplitude of $3.32\times10^{-2}$ photons $\mathrm{keV}^{-1} \mathrm{cm}^{-2} \mathrm{s}^{-1}$. We assumed that these background photons were produced close to the pulsar, and therefore accounted for absorption of the background photons while travelling to Earth. 
        
        The current likelihood computation in X-PSI is performed by marginalization over the background. Thus, the background becomes a free implicit parameter and gets degenerate with any unpulsed component of the signal\footnote{The signal has two components: a pulsed component that originates from the hot spot, and an unpulsed component that originates from the rest of star and/or the hot spot. The unpulsed component  from a hot spot may or may not be present depending on the stellar compactness, the spot location, the spot angular radius and Earth inclination to rotation axis.}. Similar to \citet{Salmi:2022}, we used a flat prior for each background variable centered around the true value with hard cuts at $\pm 3 \sigma$ to prevent the background from increasing the already large number of uncertainties \citep[see][for details]{Riley:2021,Salmi:2022}. Another reason for imposing such a constraint is that prior knowledge of the background level for real burst oscillation sources can be obtained by assuming that it is similar to the pre-burst emission. 
        
        \subsection {Priors}
       Table \ref{tab:free_params} provides a summary of the priors used for parameter inference. When choosing the parameter set for each model to generate the synthetic data, we stayed within the same prior bounds (see paragraph below for the specific case of the distance). A uniform prior is assumed by default for the majority of parameters. Following is a discussion of nontrivial prior choices.
       
       We fix the joint mass-radius prior density distribution to be flat in (M, $R_{\mathrm{eq}}$) space \citep[see][for discussion]{Riley:2018,Riley:2019}. This is done primarily to pave the way for more frequent application of posteriors in inferring the EoS. In order to account for the range of possible TBO source masses, the mass prior support is set to [1.0, 3.0] $M_{\odot}$ and the radius support to $[3r_g(1.0), 16.0]$ km with $r_g(M)=GM/c^2$ ($M$ in $M_{\odot}$ and $r_g(M)$ in km). We  set the compactness condition to $R_{\rm pole}/r_g(M) > 2.9$ and account for higher-order images expected for $R/r_g(M) < 3.52$ \citep[see][and references therein]{Bogdanov:2021} during both data generation and the inference. We imposed no restriction on the image order; thus, images are summed over until higher-images are no longer visible or visibility is truncated owing to a lack of numerical precision. During inference, the surface gravity requirement from the atmosphere model also affects the (M, $R_{\mathrm{eq}}$) prior space. In fact, samples with  $\log g \notin [13.7,14.9]$ are automatically rejected\footnote{See footenote in Sec \ref{subsubsec:atmosphere}}.
       
     When generating synthetic data that mimics the key properties of the J1814 burst oscillations, we consider distances between 0.5 and 20 kpc; but during inference, we assume a Gaussian prior with a mean equal to the distance used to create that data set, a standard deviation equal to 15\% of that distance, and a hard cutoff of $\pm 5 \sigma$.  This is done for two reasons. First, to focus on the effects of time-varying parameters and reduce the consequences of multiple-parameter degeneracy. Second, since distance constraints for TBO sources can be determined independently through methods such as photospheric radius-expansion (PRE) analysis \citep[see e.g.][]{Galloway:2020}, Gaia \citep{Gaia_Collaboration} astrometry of the donor stars \citep{Ding:2021,Moran:2022}, through pulse profile modeling of the accretion pulsations of AMPs or by modelling the burst spectra mentioned in Section \ref{sec:intro}. 
       
       To keep our simulations manageable within the restrictions of time and computational resources, we restrict the centre of the hot spot colatitude and the inclination to the stellar northern hemisphere.
     \begin{table*}
       \caption{ Parameters and their respective prior density used for inference.}
       \begin{minipage}{1.\textwidth}
        \begin{tabular}{p{5.5cm} p{8.2cm} l}
        \hline \hline
          Parameter                                 & Description                                  & Prior density \\ \hline
 
          $M$ ($M_{\odot}$)                          & Gravitational mass                          & $M\sim\mathcal{U}(1.0,3.0)$                   \\ 
          $R_{\rm eq}$ (km)                          & Equatorial radius                           &$R_{\rm eq}\sim\mathcal{U}(3r_g(1.0),16.0)$\footnote{{$r_g$: Schwarzschild gravitational radius in km}} \\
          $D$ (kpc)                                  & Distance  to Earth                          & $D\sim\mathcal{N}(d,0.15d)$\footnote{$d$: distance used to generate the synthetic data} \\
          $\cos(i)$                                  & Cosine of Earth inclination to rotation axis&$\cos(i)\sim\mathcal{U}(0.0,1.0)$                 \\
          $\phi_\mathrm{spot}$ (cycles)              & Phase of the hot region                     &$\phi_\mathrm{spot}\sim\mathcal{U}(-0.25,0.75)$   \\
          $\Theta_\mathrm{spot}$ (radian)            & Colatitude of the centre of the hot spot    &$\Theta_\mathrm{spot} \sim\mathcal{U}(0.0,\pi/2)$ \\
          $\zeta_\mathrm{spot}$ (radian)             & Angular radius of the hot spot              &$\zeta_\mathrm{spot}\sim\mathcal{U}(0.0,\pi/2)$    \\
          $\log[T_\mathrm{spot}(\mathrm{K})/1\mathrm{K}]$& Hot spot effective temperature          &$\log[T_\mathrm{spot}(\mathrm{K})/1\mathrm{K}]\sim\mathcal{U}(6.7,7.6)$\footnote{Temperature bounds set by bursting NS model described in Sec \ref{subsubsec:atmosphere}}  \\
          $\log[T_\mathrm{star}(\mathrm{K})/1\mathrm{K}]$         & Star effective temperature (K)                   &$\log[T_\mathrm{star}(\mathrm{K})/1\mathrm{K}]\sim\mathcal{U}(6.7,7.6)$           \\
          $N_{H}$   ($10^{20}\mathrm{cm}^{-2}$)              & Interstellar attenuation column density     &$N_{H} \sim \mathcal{U}(0.0,10.0)$                  \\ \hline
       \end{tabular}
       \end{minipage}
       
       \label{tab:free_params}
       \end{table*}

\subsection{Synthetic data }\label{subsec:synthetic_data}

    Bursts, in general, have changing light curves induced by temperature fluctuations as the burning progresses. Our goal in this part is to create synthetic data that mimics this type of light curve, with burst oscillations, utilizing as much of our current knowledge as possible. A burst temperature profile, which defines how the temperature evolves over time, is required to achieve this goal when utilizing X-PSI.
    
    
   Given the uncertainty associated with burning physics, we were unable to obtain from the literature a simple analytical expression for such a temperature. \citet{Bhattacharyya:2006b} proposed a functional (exponential) form for the temperature, but  using their formula requires a few additional free parameters. \citet{Galloway:2020} have since shown that a power-law model, or a Gaussian plus a power-law model, yields a more accurate fit in the tail than an exponential decay. This implies additional customizable settings. In order to reduce the number of free parameters associated with the temperature profile, we instead used a previously established temperature profile acquired through simulation.

\begin{figure}
    \centering
    \includegraphics[width=1\columnwidth]{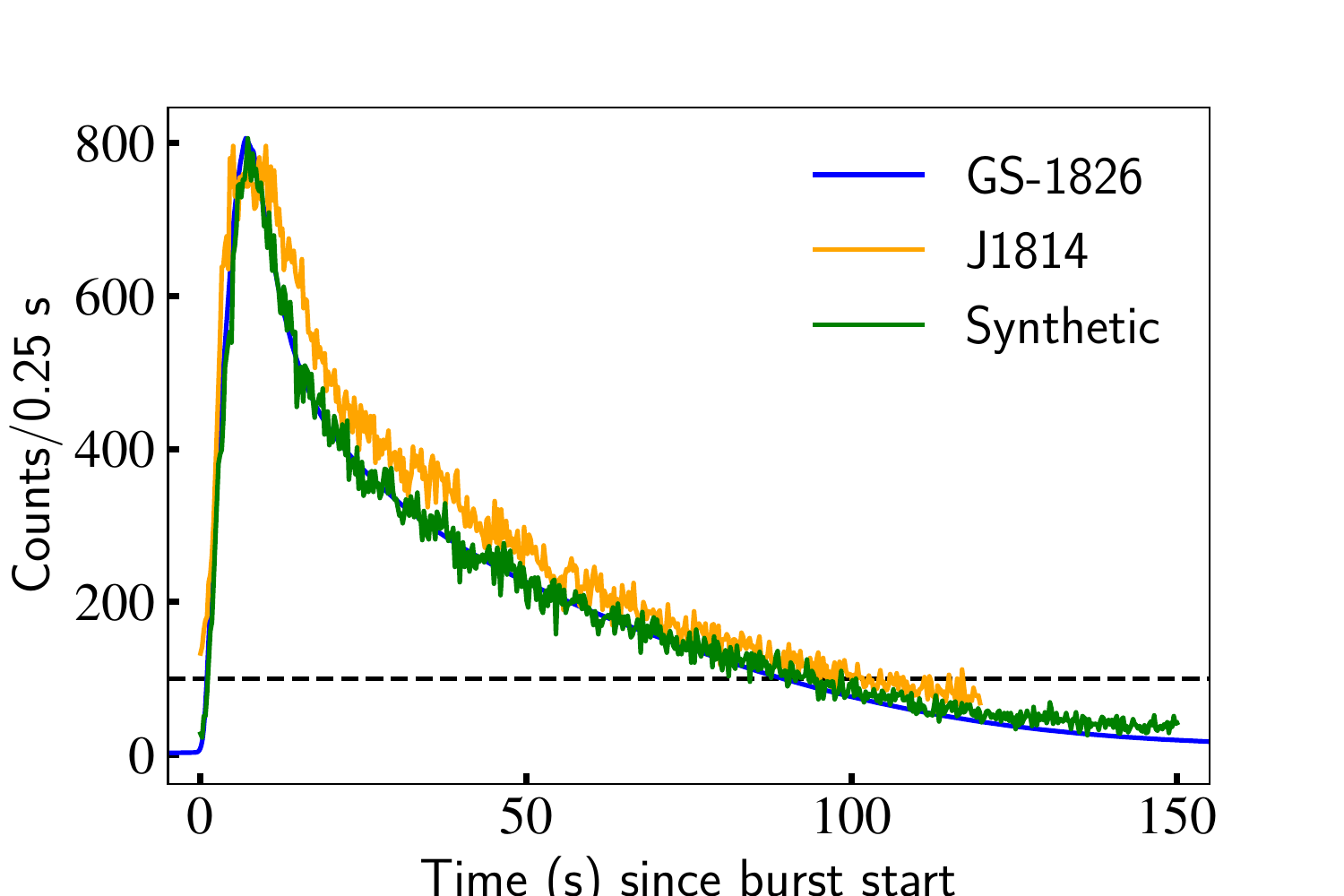}
    \caption{Synthetic burst light curve produced using parameter set 2 of model ST-$\bar{S}\tilde{H}_{T,R}$, compared to an actual burst of J1814 (burst 26) and the average light curve of GS 1826$-$24  normalised by the synthetic burst maximum counts. The black dashed horizontal line is the background threshold above which the \textit{chopped burst} is considered.}
    \label{fig:burst}
    \end{figure}

    Comparing the light curves of GS 1826$-$24  and J1814 \citep[see][for light curves \footnote{\url{https://burst.sci.monash.edu/}}]{Galloway:2020} suggested that  GS 1826$-$24 is a reasonable mimic for J1814 (see Figure \ref{fig:burst}). The burst light curves of both sources have a short rise time and a slow exponential decay. This is consistent with a helium flash followed by a $rp$-process burning of hydrogen \citep{Schatz:2001} and implies similar burning physics. Therefore, to obtain the temperature profile for the analysis, we used the simulated temperature profiles of GS 1826$-$24 obtained using the stellar and binary evolution code MESA \citep[Modules for Experiments in Stellar Astrophysics;][]{Paxton:2011},  then averaged all simulated bursts' temperatures in the ma6 model sequence \citep{Meisel:2018rsy} apart from the first and the last. We exclude these two since the initial burst is typically more energetic than the average burst, and the last burst was incomplete. We note that model ma6 has the same chemical composition as that of the atmosphere model used throughout the modeling. We denote by $T_\mathrm{ma6}(t)$ the final temperature profile obtained, where $t$ is the time.

\begin{figure}
    \centering
    \includegraphics[width=\columnwidth]{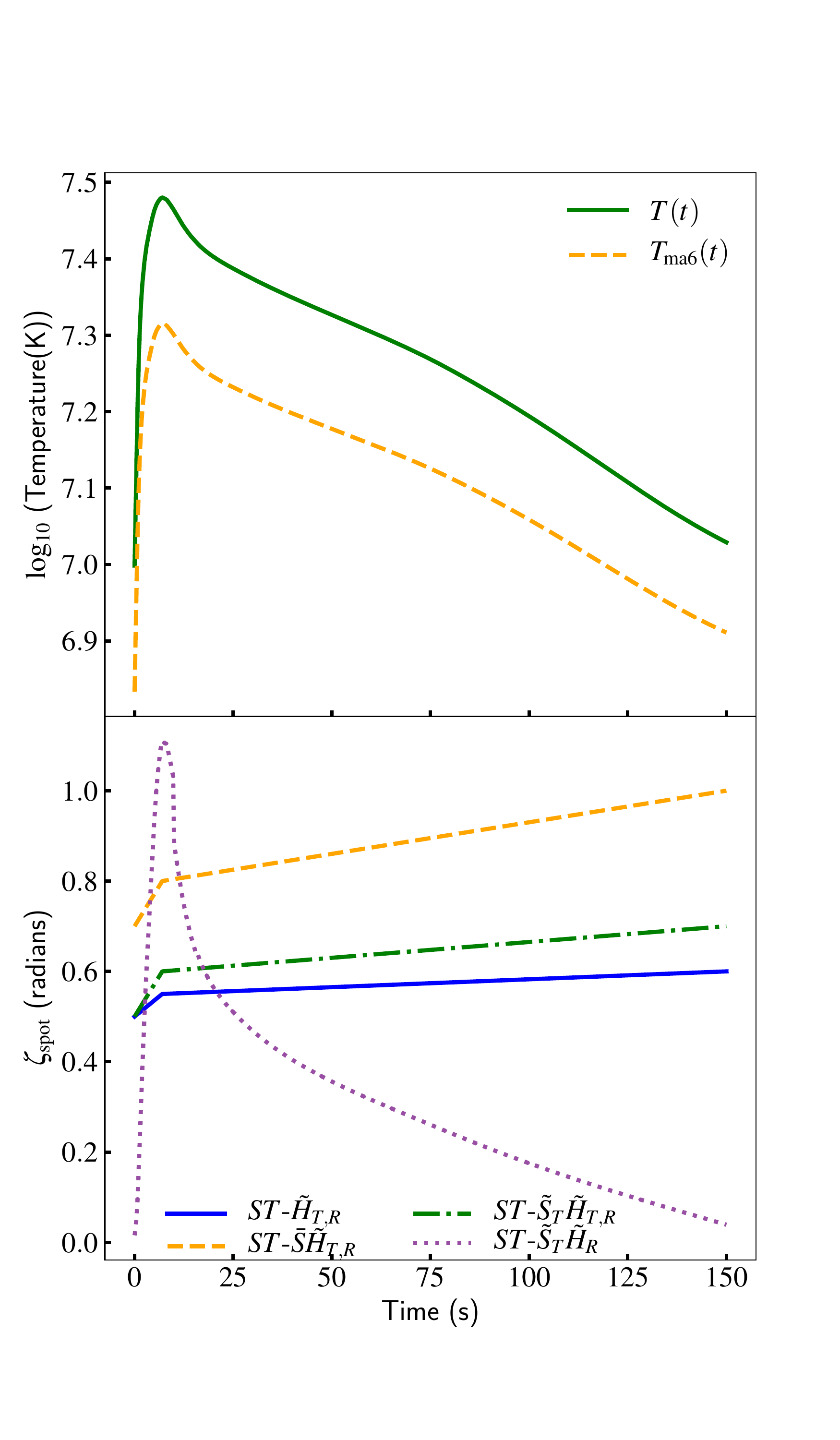}
    \caption{Temperature (top) and angular radius (bottom) profile. $T(t)$ is the hot spot temperature profile ($T_\mathrm{spot}(t)$) for models: ST-$\tilde{S}_{T}\tilde{H}_{T}$ and ST-$\tilde{S}_{T}\tilde{H}_{T,R}$. $T_\mathrm{ma6}(t)$ represents the hot spot temperature profile ($T_\mathrm{spot}(t)$) for models: ST-$\tilde{H}_{T}$, ST-$\tilde{H}_{T,R}$, ST-$\bar{S}\tilde{H}_{T}$ and ST-$\bar{S}\tilde{H}_{T,R}$ and the star temperature ($T_\mathrm{star}(t)$) for models: ST-$\tilde{S}_{T}\tilde{H}_{T}$, ST-$\tilde{S}_{T}\tilde{H}_{T,R}$, ST-$\tilde{S}_{T}\bar{H}$ and  ST-$\tilde{S}_{T}\tilde{H}_{R}$}
    \label{fig:profils}
    \end{figure}

    Since some models require two temperature profiles, one for the hot spot and one for the star, we manually created a second profile and named it $T(t)$. We also crafted an angular radius profile denoted $\zeta_\mathrm{spot}(t)$ for every model that required it. Both $T$(t) and $\zeta_\mathrm{spot}(t)$ are shown in Figure \ref{fig:profils} and were adjusted for each model such that the resulting light curve is comparable to J1814, the number of photons is approximately 10$^5$, and the FA is approximately 10\%.
    
    For each model, we set the temperatures as follows:
    \begin{itemize}
    \item $T_\mathrm{spot}$(t)=$T_\mathrm{ma6}(t)$ for ST-$\tilde{H}_{T}$, ST-$\tilde{H}_{T,R}$, ST-$\bar{S}\tilde{H}_{T}$ and ST-$\bar{S}\tilde{H}_{T,R}$;
    \item $T_\mathrm{spot}$(t)=$T(t)$ and $T_\mathrm{star}(t)$=$T_\mathrm{ma6}$(t) for  ST-$\tilde{S}_{T}\tilde{H}_{T}$ and ST-$\tilde{S}_{T}\tilde{H}_{T,R}$;
    \item $T_\mathrm{star}(t)$=$T_\mathrm{ma6}(t)$ for  ST-$\tilde{S}_{T}\bar{H}$ and  ST-$\tilde{S}_{T}\tilde{H}_{R}$.
    \end{itemize}

     We discretized all temperatures and $\zeta_{\mathrm{spot}}(t)$ into arbitrary 0.25 s time intervals.  We then generated 0.25 s  exposure time pulses  for a total of 150 seconds and summed those pulses to obtain the desired pulse and light curve (see Figure \ref{fig:burst}). Each 0.25-second pulse is the sum of the expected counts from the model and a Poisson fluctuation. We used the random number generator of the GNU Scientific Library \citep[GSL;][]{Gough:2009} to add this Poisson fluctuation. To ensure that a different noise is generated for each time segment, the seed environment variables of GSL were changed for each segment. Snapshots of the pulses corresponding to the synthetic light curve of Figure \ref{fig:burst} are shown in Figure \ref{fig:pulse} of Appendix \ref{sec:appendix}.
     
     For this paper, we do not allow the background level to vary during the burst. Hence, we set the expected bolometric background counts to 20 in each 0.25 s time bin. This yields  background counts equivalent to 10\% of the total photon counts as observed in J1814 bursts. To avoid having the background dominate the data, we only consider the portion of the synthetic burst from the point where the counts during the rise for the first time exceed five times the injected background counts, to the point in the decay where the counts are for the last time at least five times the injected background counts. We refer to this part of the synthetic burst as the \textit{chopped burst} and this is used for parameter inference.
     
     A typical burst of J1814 has 10$^5$ counts and 10\% rms FA. Therefore, a set of parameters is selected if and only if the \textit{chopped burst} shape is comparable to that of the averaged of GS 1826$-$24 light curves, produces 10$^5$ counts and 10\% rms FA, and if the rms FA is relatively constant during the \textit{chopped burst}. To guarantee that the final requirement is met, we segment the \textit{chopped burst}  in bins of 5000  photons and compute the rms FA for each bin. For each model, we retained three parameter sets and summarize them in Table \ref{tab:params_var} of Appendix \ref{sec:appendix}.

     Since a  typical burst of J1814 has 10$^5$ counts, it is possible, in principle, to get tighter parameter constraints by combining more than one burst from the same source. To explore if the parameter recovery is affected when multiple bursts are combined, we produced  100 synthetic bursts for each selected parameter set and merged the first 10 and 100 synthetic bursts to form a \textit{mega} burst of respectively $10^6$ and  $10^7$ photons. For each burst in the 100 synthetic bursts sequence, only the GSL seeds were allowed to vary during and between bursts. 
     The model parameters, such as the hot spot location, angular radius, temperature profile, etc., were considered fixed from one burst to another\footnote{{This is an ideal scenario. For most sources, hot spot location, angular radius, temperature profile, etc. are expected to vary from burst to burst \citep[see e.g.][]{Muno:2002, Galloway:2020}. This is something that could and should be investigated in more detail in future work. However, given the similarities in some bursts of 1814, merging those bursts assuming that the temperatures and/or radii profiles are identical is a reasonable preliminary approach.}}.
     
     We refer to all data created using a time dependent temperature (and angular radius, depending on the model) as the \textit{variability data set}.

    Parameter degeneracy in pulse profile modeling  is well-known and can be difficult to overcome \citep{Lo:2013ava}. Prior to submitting the variability data set to the simulation pipeline, we must ensure that the method by which each burst was generated did not introduce biases that could lead  to inaccurate parameter recovery. Moreover, we use nested sampling package \texttt{MultiNest} to estimate the posterior probability density function (PDF) for each parameter. \texttt{MultiNest} samples the prior hyper-space and approximates the true PDF and depending on the nature of the problem, its settings must be adjusted for optimal performance.  To do so, we created a \textit{calibration data set} where for each model, the time dependent variable has been set to a fixed value such that the photon counts and the FA remain as close as possible to their the \textit{variability data set} counterpart (see Table \ref{tab:params_const} in Appendix \ref{sec:appendix}). The exposure time remains as it would be in for chopped burst. For the  $10^6$ and  $10^7$ photon count bursts, only the exposure times were extended to yield the desired counts.

   In summary, we have generated two data sets. The first data set, referred to as the  \textit{calibration data set}, contains no temporal variation. The hot spot temperature, angular radius, and stellar temperature (if applicable to that model) were set to remain constant. For the \textit{variability data set}, however, the temperatures and/or angular radius were time-dependent variables conditional on the model. We utilized both data sets and conducted inference runs based on a model that was used to generate the \textit{calibration data set}, in which variability is  eliminated. Our findings are presented below.

\section{Results}\label{sec:result}

    We attempted to recover the parameters by conducting an inference run on each parameter set for each model. Given three parameter sets per model, eight models per data set, two data sets, and three cases with a different number of count we performed a total of  one hundred forty-four inference runs\footnote{The synthetic data and the inference run outputs are available \href{http://10.5281/zenodo.7665653}{here} \citep{zenodo_paper}}. We began by examining the quality of data recovery per data set. Then we examined each model's performance with respect to the \textit{variability data set}.

     To estimate the quality of parameter recovery, we make use of the PP-plot\footnote{The PP-plot has on the x-axis the expected cumulative probability and on the y-axis the observed cumulative probability calculated from the data.} which is a tool used to compare two distributions and assessing their similarities or differences. 
     This  method  is commonly used in gravitational-wave studies to evaluate the robustness of inference processes \citep{Sidery:2014,Veitch:2015,Berry:2015,Gair:2015,Pankow:2015,Biwer:2019,Romero-Shaw:2020,Cornish:2021,You:2022,Mozzon:2022}. 
    Assuming the inference process has no inherent bias, the injected (or true) value of each parameter must fall inside the x\% credible interval x\% of the time \citep{cook_2006}. This implies that the fraction of the recovered parameter within a given credible interval must be uniformly distributed. Hence, the cumulative fraction of the recovered parameter should sit on the diagonal of the PP-plot, with occasional fluctuations due to the limited sample size. We therefore perform a Kolmogorov–Smirnov test (KS-test) to determine the extent to which the cumulative fraction of recovered parameters deviates from the uniform distribution.

    We introduce a wrongness parameter, in addition to the PP-plots, to cross-check the quality of parameter recovery. We define wrongness = $({\mathrm{X}^\mathrm{inf}}$-$\mathrm{X}^{\mathrm{inj}})/\Delta \mathrm{X}^{\mathrm{inf}}$, where  X$^{\mathrm{inf}}$ and X$^{\mathrm{inj}}$ are respectively the inferred and the injected values (median of the posterior) of the parameter X. $\Delta \mathrm{X}^{\mathrm{inf}}$ is the 68\% credible interval recovered from sampling. Alternatively stated, wrongness is the difference between the inferred and injected values in units of 68\% credible interval (defined around the median as in the posterior plots), and would correspond to the standard deviation if the posteriors were Gaussians. A perfect parameter recovery would result in a wrongness of zero. We expect, however, that, due to statistical fluctuations and systematic error during sampling, the wrongness of all runs for a particular parameter will cluster close to zero.

    In the following, we focus mainly on the space-time parameters: compactness, radius, and mass; the primary parameters of interest for dense matter studies. Alongside the space-time parameters, we also inferred the geometrical parameters. Some are recovered more effectively than their EoS equivalents, while others are not. However, we provide fewer details of the geometric parameter inference since they are not the primary focus of this investigation.

    \begin{figure*}
    \hspace*{-1cm} 
    \centering
    \includegraphics[width=2.5\columnwidth]{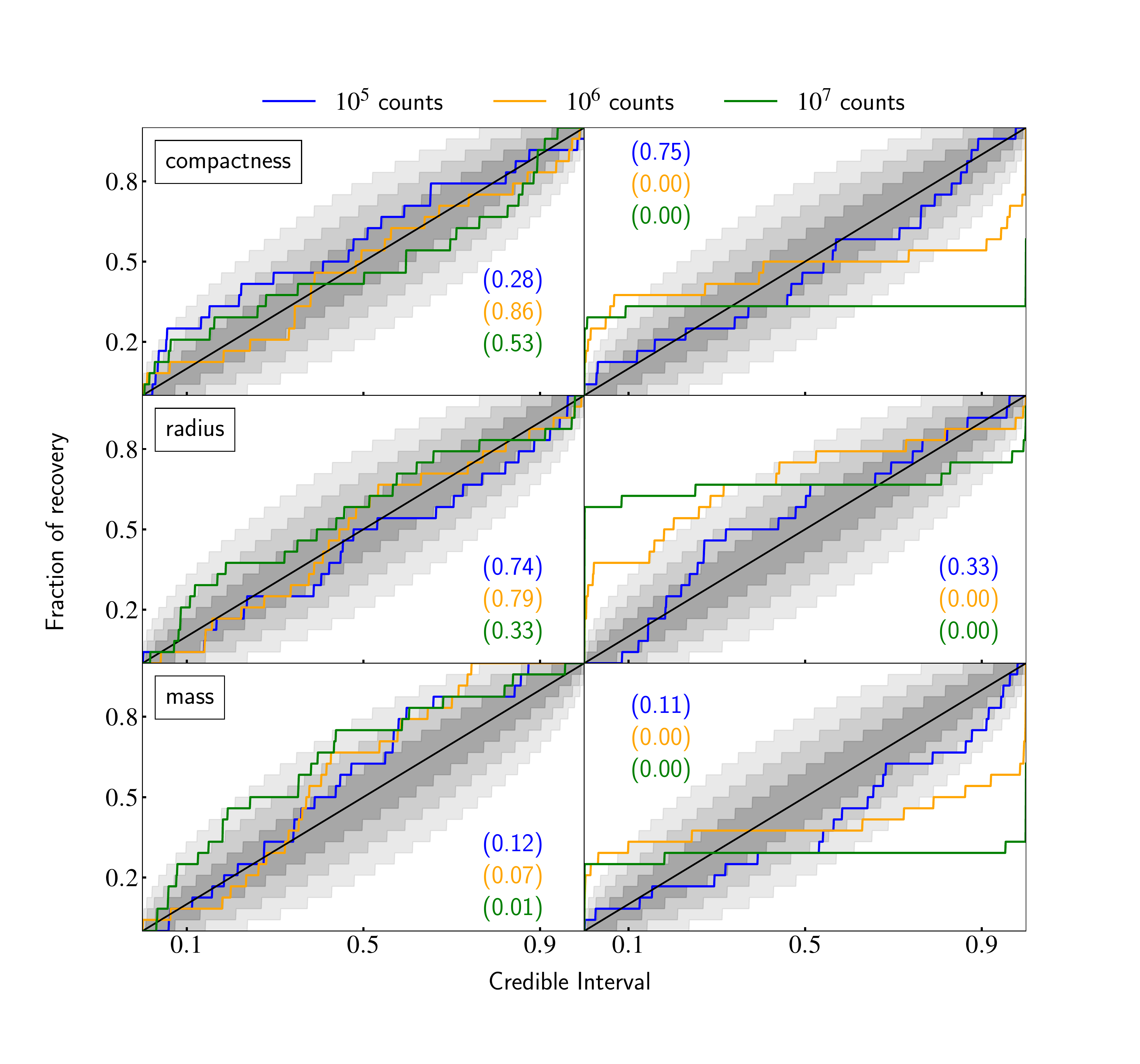}
    \caption{Cumulative fraction of parameter values (used to generate the data) recovered within a credible interval as a function of the credible intervals  (from the lower tail of the posterior distribution) for the \textit{calibration data set} (left) and \textit{variability data set} (right). The grey regions, from most opaque to least, depict the cumulative 1, 2, and 3$\sigma$ confidence intervals.  The black solid line represents the expectation for unbiased parameter estimates. The  blue, orange, and green curves correspond respectively to 10$^5$, 10$^6$, 10$^7$ counts with their associated \textit{p}-values obtained from the KS test values in parenthesis.}
    \label{fig:pp-plot}
    \end{figure*}

    \subsection{Calibration data set}

    We gauge the validity of our \texttt{MultiNest} settings and check for any bias by performing parameter recovery using the \textit{calibration data set}.
    
   The PP-plot in Figure \ref{fig:pp-plot} shows the cumulative fraction of parameter values recovered within a credible interval\footnote{The x\% credible interval in the PP-plot is the interval between the lower tail of the distribution and the point below which x\% of the samples are located. This is different than that used in the posterior 1D plots.}
   as a function of the credible intervals for the \textit{calibration data set} (left).  The top, middle, and bottom panels are the recovery fraction for the compactness, radius, and mass respectively. Each color represents a given number of counts, as indicated on top of the figure. The numbers in parenthesis are the \textit{p}-values from the KS-test for each number of counts in the corresponding color. The gray shaded regions are the $\pm$1, 2 and 3$\sigma$ expected uncertainty regions corresponding to a sample size of 24 (all the runs corresponding to one number of counts)  and are derived from the binomial quantile \citep{Cameron:2011}. 
   
\begin{figure}
    \centering
    \includegraphics[width=\columnwidth]{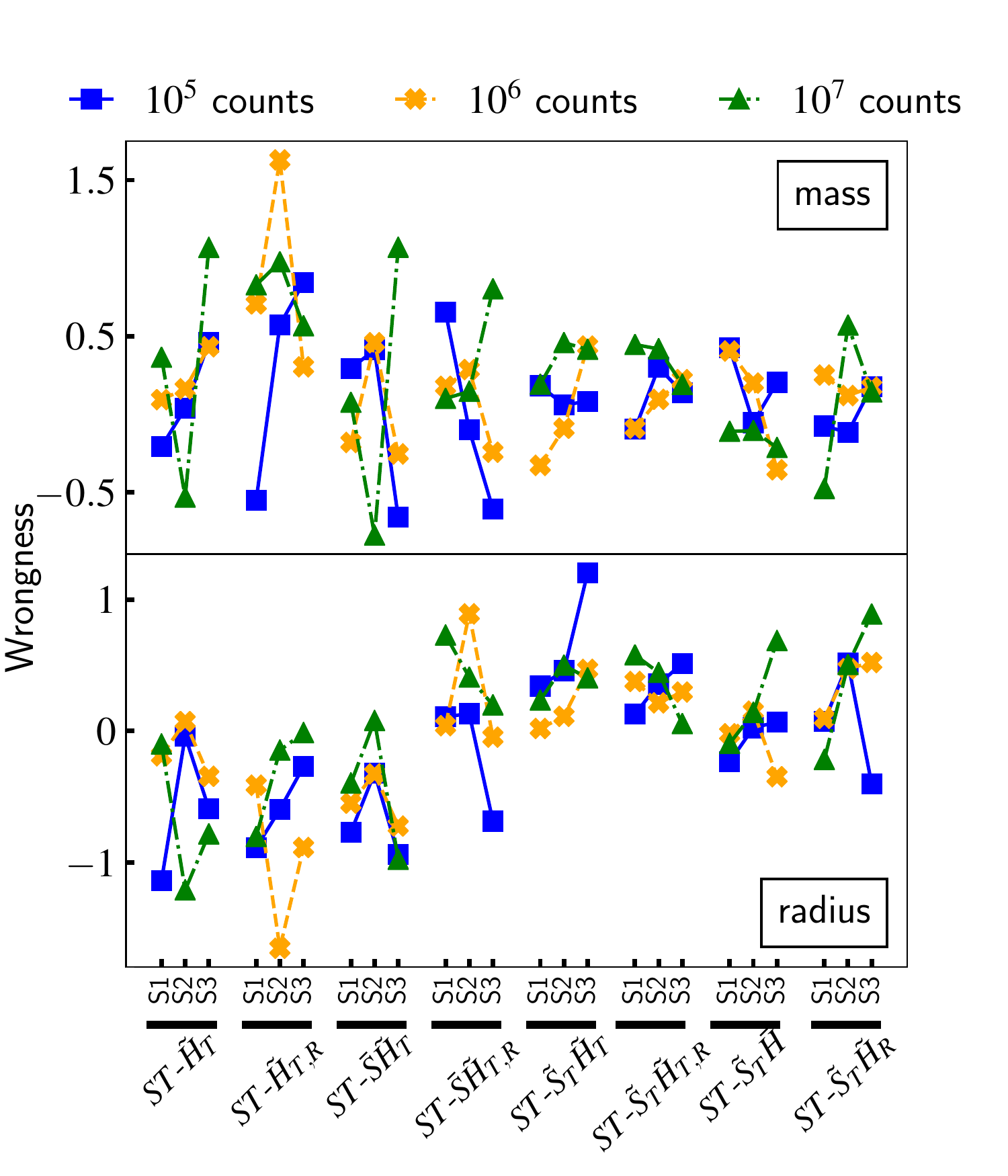}
    \caption{Wrongness, using the \textit{calibration data set} mass and radius calibration data set.  The Wrongness = $(X^{\mathrm{inf}}-X^{\mathrm{inj}})/\Delta X^{\mathrm{inf}}$, is a measure of how far off the inferred value ($X^{\mathrm{inf}}$) is from the injected value ($X^{\mathrm{inj}})$ in units of 68\% credible interval  (around the median); X $\in \{ M, R_{\mathrm{eq}}\}$. On the x-axis are the different models and their parameter sets with three different parameter sets S1, S2, and S3 for each model.}
    \label{fig:wrongness_calibration}
    \end{figure}

   For this data set, the recovered compactness and radius are extremely close to what is expected for all the counts and yield \textit{p}-values $>$ 0.28. The mass is also recovered effectively for 10$^5$ and 10$^6$ counts, but is somewhat worse for 10$^7$ counts. The curve always lies above the expected recovery, which implies that the majority of injected masses are below the 0.5 probability mass of the recovered posterior. A better recovery might be obtained by raising \texttt{MultiNest}'s live points and/or lowering the sampling efficiency. However the cumulative fraction of parameter values recovered within a credible interval remains inside the $\pm3\sigma$ expected region. In Figure \ref{fig:wrongness_calibration}, we plot both mass (top panel) and radius (bottom) wrongness for all 24 runs. For the radius, the mean wrongness is -0.12, -0.07, and 0.04 for respectively 10$^5$, 10$^6$, and 10$^7$ counts. The wrongness is centered near zero and decreases as the number of photons increases, which implies radius is well recovered. For the mass, the mean wrongness is 0.09, 0.19, and 0.27 for respectively  10$^5$, 10$^6$, and 10$^7$ counts. The average wrongness tends to increase as the data quality improves which indicates a small bias toward high inferred masses. Nevertheless, the maximum wrongness for the mass is about 1.6, which indicates that the recovered mass is not far from the injected value. 
   
\begin{figure}
    \centering
    \includegraphics[width=1.\columnwidth]{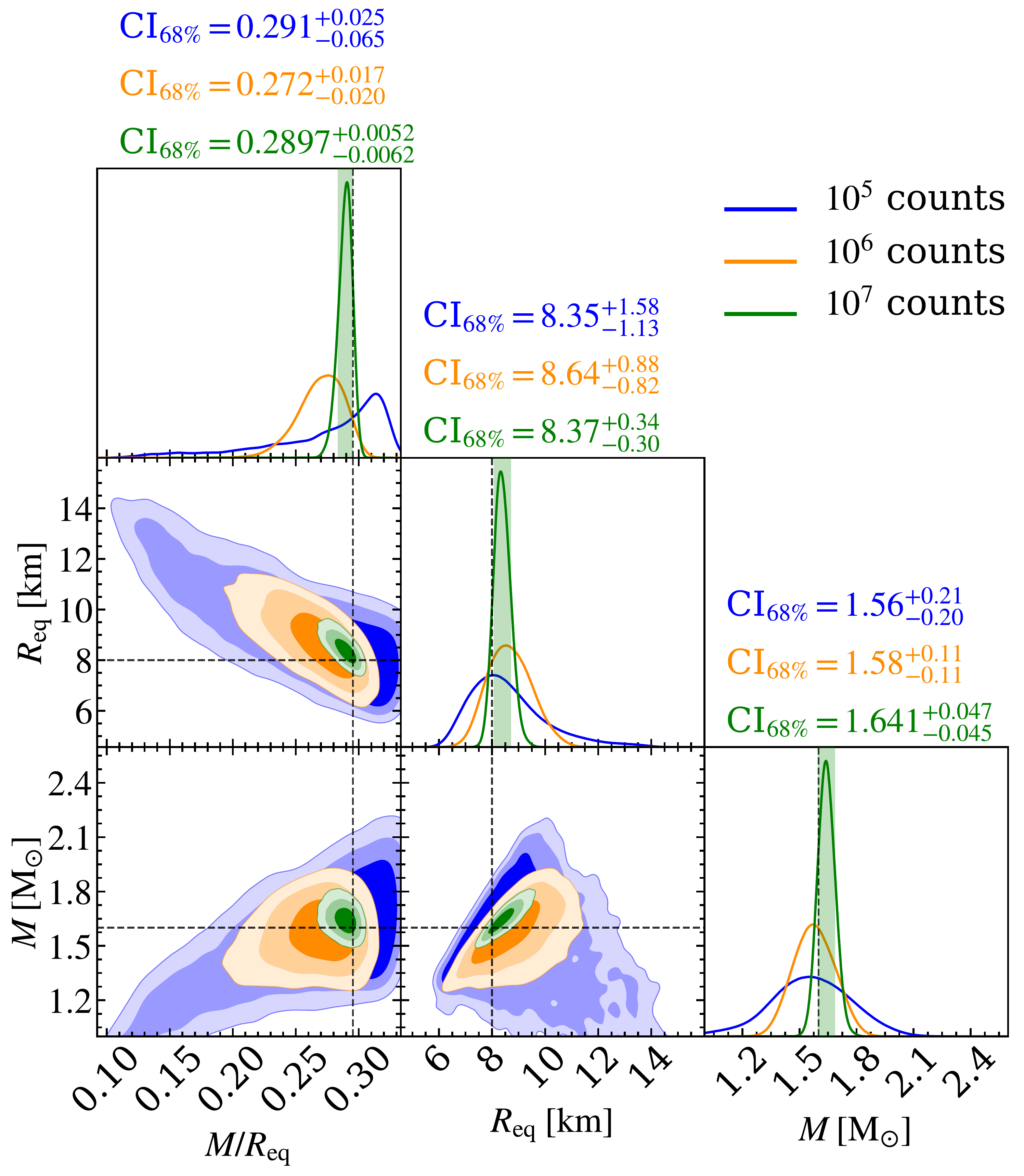}
    \caption{Example of posterior plots obtained using data corresponding to set 1, model ST-$\tilde{S}_{T}\tilde{H}_{T,R}$ selected from the  \textit{calibration data set}. The dashed black lines are the values injected to generate the synthetic data. The green vertical band shows a 68\%  posterior density credible interval symmetric in posterior mass around the median corresponding to 10$^{7}$ photon counts. The 2D posterior distributions show the 68\%, 95\% and 99\%  posterior credible region, respectively, from most opaque to least.}
    \label{fig:posterior_calibration}
    \end{figure}

     
   Another method of examining parameter recovery is by looking at posterior plots. In Figure \ref{fig:posterior_calibration}, we show a graphical representation of how the posteriors behave as the quality of the data improves (as the number of counts increases). This posterior plot corresponds to parameter set 1 of the ST-$\tilde{S}_{T}\tilde{H}_{T,R}$  model \textit{calibration data set} counterpart. As the number of counts increases, the credible intervals of the compactness, the radius, and the mass tighten up and become narrower around the injected (true) value. This is the expected behavior of parameter recovery when there is no bias in the inference process.

    The PP plots, the posterior plots, and the wrongness plots suggest the absence of significant biases in our inferences of compactness, mass and radius for the \textit{calibration data set}. Therefore, there is no evident bias introduced by the data creation method. Thus, we apply the same \texttt{MultiNest} settings to the \textit{variability data set}.


     \begin{figure}
    \centering
    \includegraphics[width=\columnwidth]{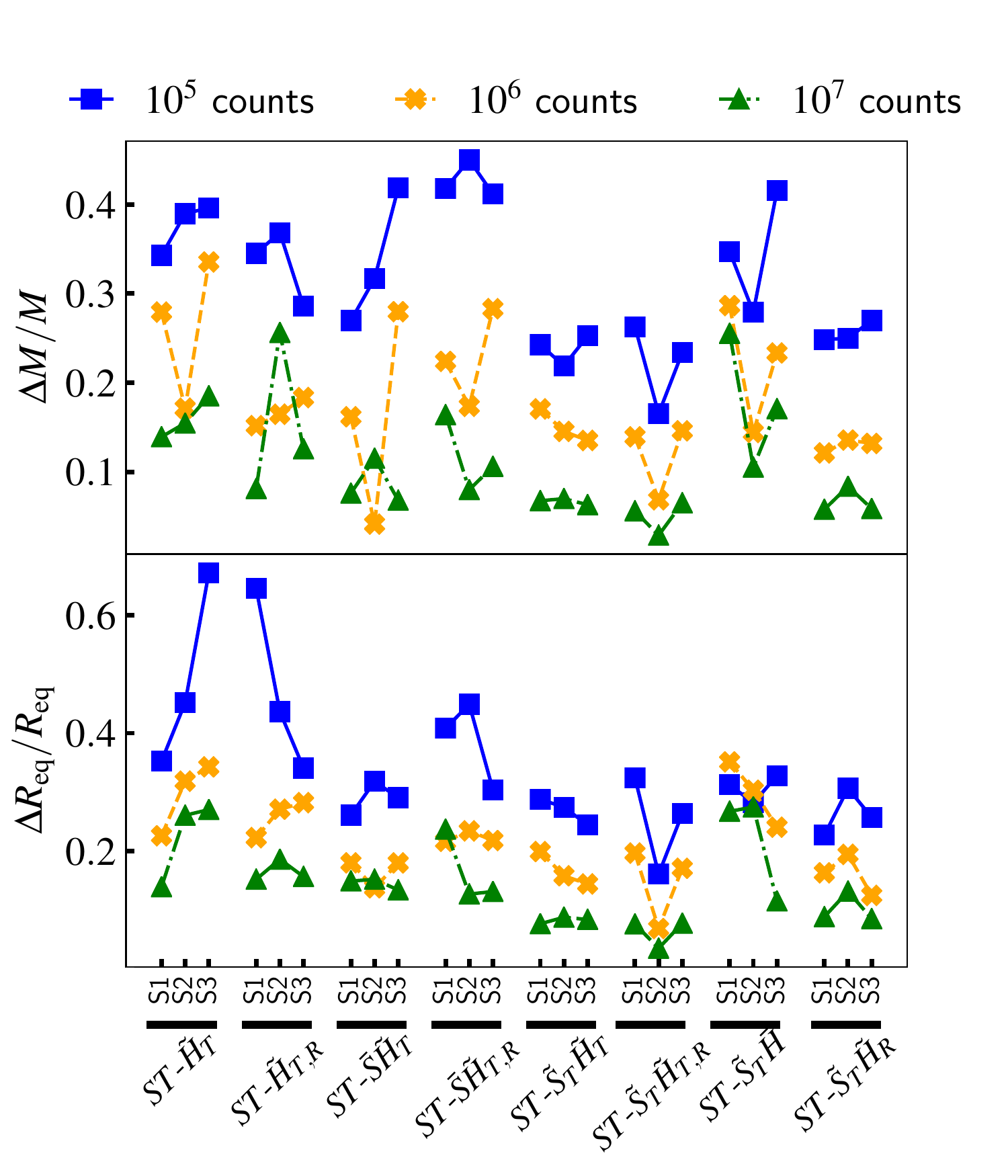}
    \caption{ Mass and radius uncertainty at 68\% credible interval for the \textit{calibration data set}. On the x-axis are the different models and their parameter sets with  SX the parameter set X of the model. The models indicated  on the x-axis correspond to their variability counterparts, upon which synthetic data were generated using the method outlined in Section \ref{subsec:synthetic_data}}
    \label{fig:mass_radius_uncertainty}
    \end{figure}

    Using the \textit{calibration data set}, we examined the uncertainties in the radius and mass. The uncertainty is defined as $\Delta \mathrm{X/X}$ where $\Delta \mathrm{X}$  is the  68\% credible interval of the inferred parameter X. This is shown on Figure \ref{fig:mass_radius_uncertainty} for each parameter set of each model. In general, the uncertainties decrease as more counts are collected, as expected. The radius has larger fractional uncertainties than the mass. This provides more justification for why the radius recovery seems better than that of the mass from the PP-plots.

    \subsection{Variability data set}

    In the right column of Figure \ref{fig:pp-plot}, we present  the cumulative fraction of parameter values recovered inside a credible interval as a function of the credible intervals for the \textit{variability data set} for the compactness (top), radius (middle), and mass (bottom). When time variability is overlooked during inference, parameter recovery for 10$^5$ counts is effective. At this low data quality regime, the PP-plots remain within the expected 3$\sigma$ credible interval and produce statistically acceptable \textit{p}-values for all EoS parameters. As anticipated, the parameter recovery is lower than the \textit{calibration data set} counterpart for mass and radius. For compactness, however, the recovery is better, primarily due to statistical fluctuation and our relatively small sample size.
    
    The PP-plots diverge, however, from the expected behavior as the data quality improves (10$^6$ and 10$^7$ counts). Both the 10$^6$ and 10$^7$ counts graphs exceed the $\pm3\sigma$ expected region, with the 10$^7$ count case performance being much worse. More injected values than expected are found at lower credible intervals, for both 10$^6$ and 10$^7$ counts, but fewer injected values than expected are found at higher credible intervals. This implies an excessively narrow posterior and substantial biases. 
   
\begin{figure}
    \centering
    \includegraphics[width=1.\columnwidth]{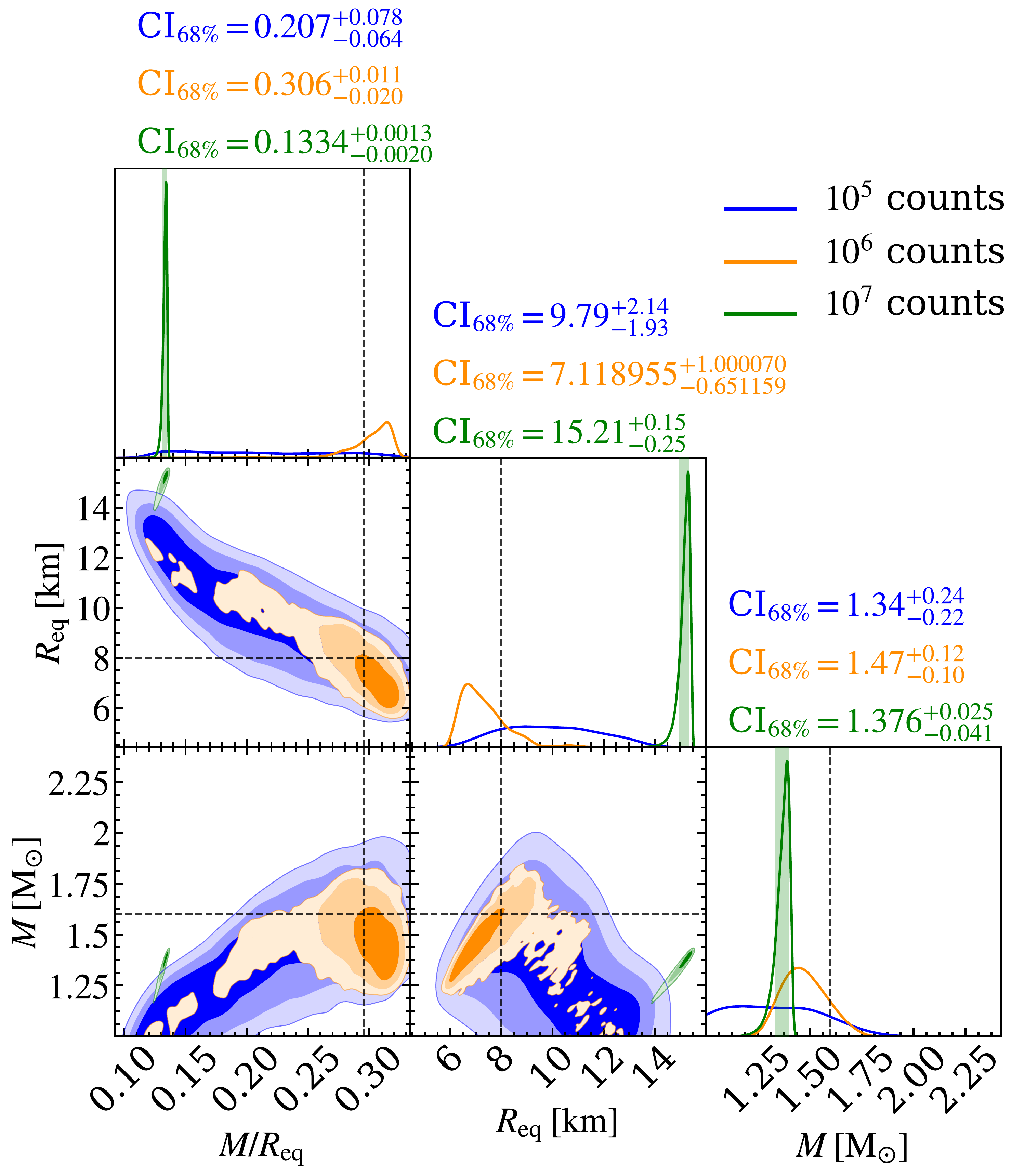}
    \caption{Example of posterior plots obtained using data corresponding to set 1, model ST-$\tilde{S}_{T}\tilde{H}_{T,R}$ selected from the  \textit{variability data set}. The dashed black lines are the values injected to generate the synthetic data. The green vertical band shows a 68\%  posterior density credible interval corresponding to 10$^{7}$ photon counts. The 2D posterior distributions shows the 68\%, 95\% and 99\%  highest posterior credible region, respectively, from most opaque to least.}
    \label{fig:posterior_variabilty}
    \end{figure}

    To emphasize this point, we show  on the right panel of Figure \ref{fig:posterior_variabilty} the posterior of the compactness, mass, and radius. This figure shows that for 10$^5$ counts, even when temporal variability is overlooked, the posteriors are as broad as their calibration counterpart and encompass nearly the whole parameter space. In a high data quality regime, especially for 10$^7$ counts, the posterior is compressed to a small region of the parameter space. This is because most of the parameter vectors in the prior space fail to yield good likelihood values, resulting in the most of samples clustering in smaller regions with relatively higher likelihood. This is the consequence of using an inappropriate model to infer the parameters. 

\begin{figure}
    \centering
    \includegraphics[width=1.\columnwidth]{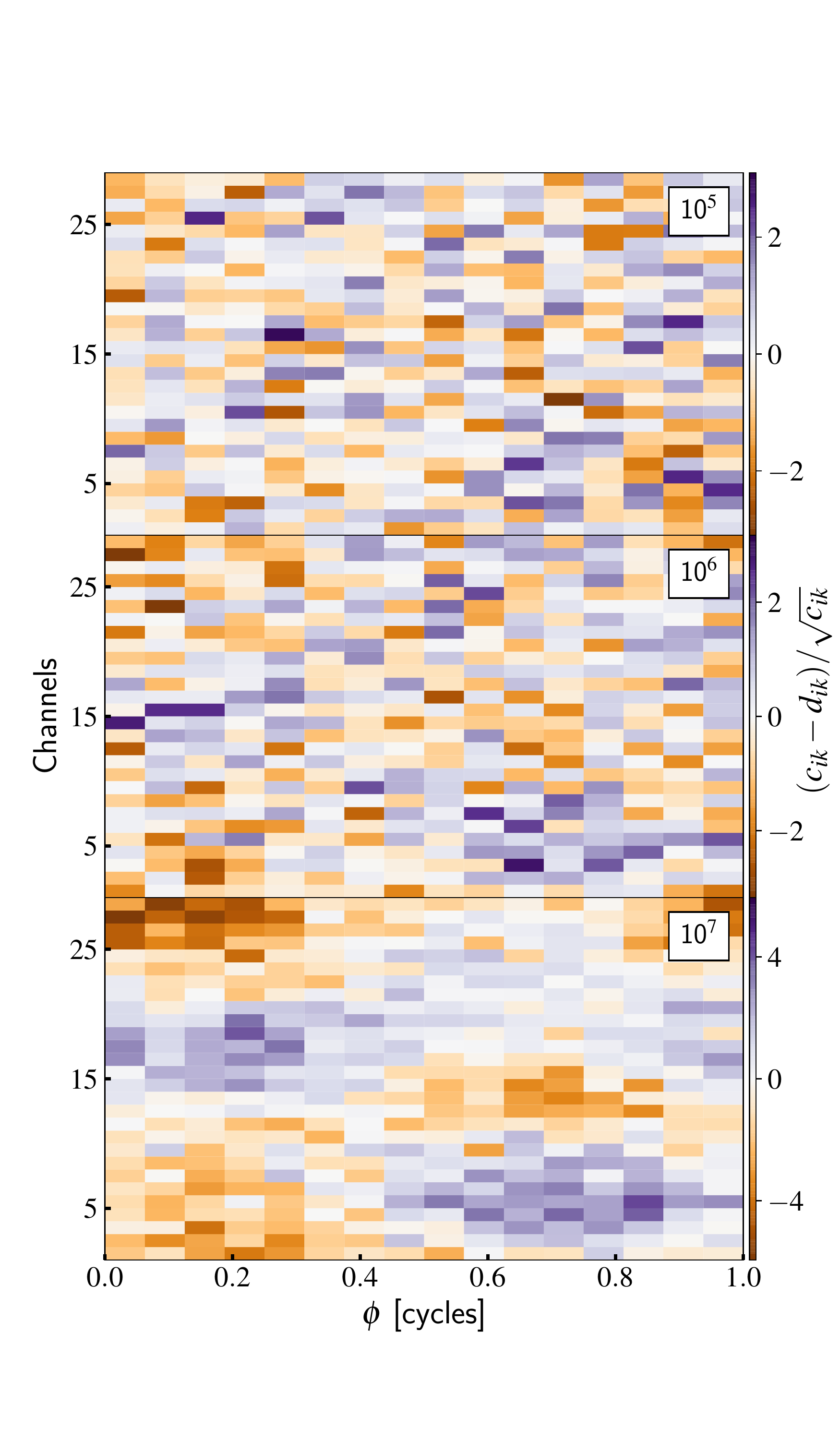}
    \caption{Residuals in instrument energy channels as a function of rotational phase and energy channel, indicating the difference between expected counts from the maximum likelihood solution and actual data counts normalized by the expected counts. $c_{ik}$ and $d_{ik}$ denote respectively the model counts and the data counts in the $i^{th}$ rotation phase and  $k^{th}$ energy channel.}
    \label{fig:residuals}
    \end{figure}

    To provide more evidence of such poor model performance, Figure \ref{fig:residuals} shows the residuals corresponding to the posterior of Figure \ref{fig:posterior_variabilty}. The residuals were computed using the maximum likelihood parameter vector in each case. The residuals for 10$^5$ counts appear normal. Due to the rather poor quality of the data, the majority of the parameters fit the data relatively easily. However, at higher counts, particularly at 10$^7$ counts, discernible patterns emerge. The model generates an excess of counts in certain regions highlighted in blue and a deficiency in others (in red), with clusters that are large enough to indicate a significant deviation from randomness.  These clusters are indicative that the fitting under this data regime is inadequate.This is not unexpected, given that the model used to create the data differs from the model used for inference. 

\begin{figure}
    \centering
    \includegraphics[width=1\columnwidth]{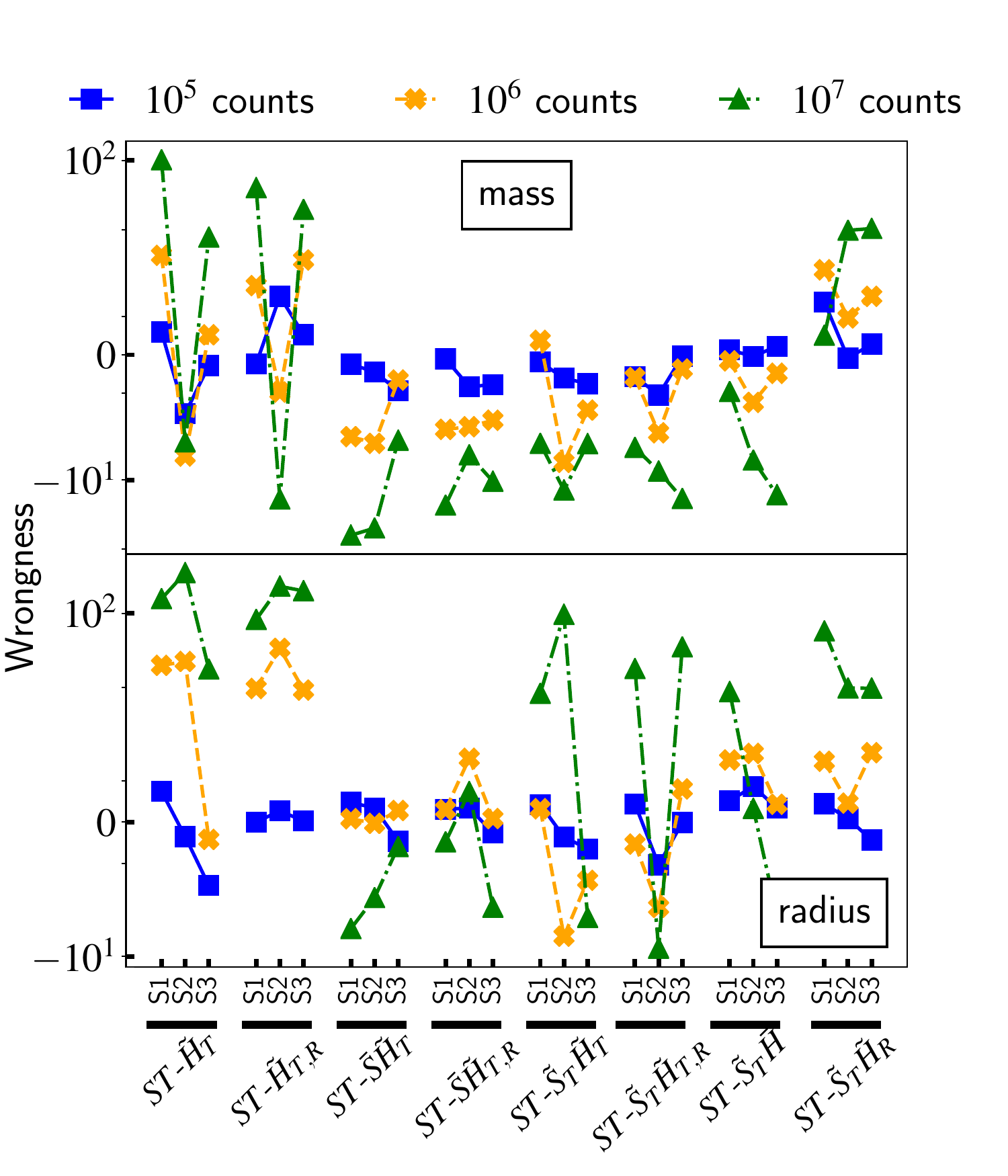}
    \caption{ Wrongness using the \textit{calibration data set} mass and radius calibration data set.  The Wrongness = $(X^{\mathrm{inf}}-X^{\mathrm{inj}})/\Delta X^{\mathrm{inf}}$, is a measure of how far off is the inferred value ($X^{\mathrm{inf}}$) from the injected value ($X^{\mathrm{inj}})$ in units of 68\% credible interval around the median) ; X $\in \{ M, R_{\mathrm{eq}}\}$. On the x-axis are the different models and their parameter sets with three different parameter sets S1, S2, and S3 for each model.}
    \label{fig:wrongness_variability}
    \end{figure}

    As we did for the \textit{calibration data set}, we also plot the wrongness for both the mass and radius in Figure \ref{fig:wrongness_variability}. Unlike in the \textit{calibration data set} case, the wrongness moves away from zero when the data improves and yields values up to 100 for some models. This provides evidence that the posteriors have been overly constrained away from the injected values. Only a few models yield acceptable wrongness values at 10$^7$ counts. To check whether data generated under any specific model performs better than others, we show the PP-plot in Figure \ref{fig:compactness_per_model} for the compactness (see Figure \ref{fig:radius_per_model} in Appendix \ref{sec:appendix} for the radius and mass). The top and bottom panels correspond respectively to 10$^6$ and 10$^7$ counts. On top of the figure, each model and its corresponding \textit{p}-values (left in parenthesis: 10$^6$ counts, right in parenthesis: 10$^7$ counts) are highlighted. Given the small sample size per model (three inference runs), it is difficult to draw meaningful conclusions regarding the quality of the recovery based solely on this PP-plots. However, we display the performance of each model as a visual indication. Despite the substantial uncertainties, most of the models fall outside the expected 3 $\sigma$ region and none of the models seems to perform well.

      
    \begin{figure}
    \centering
    \includegraphics[width=\columnwidth]{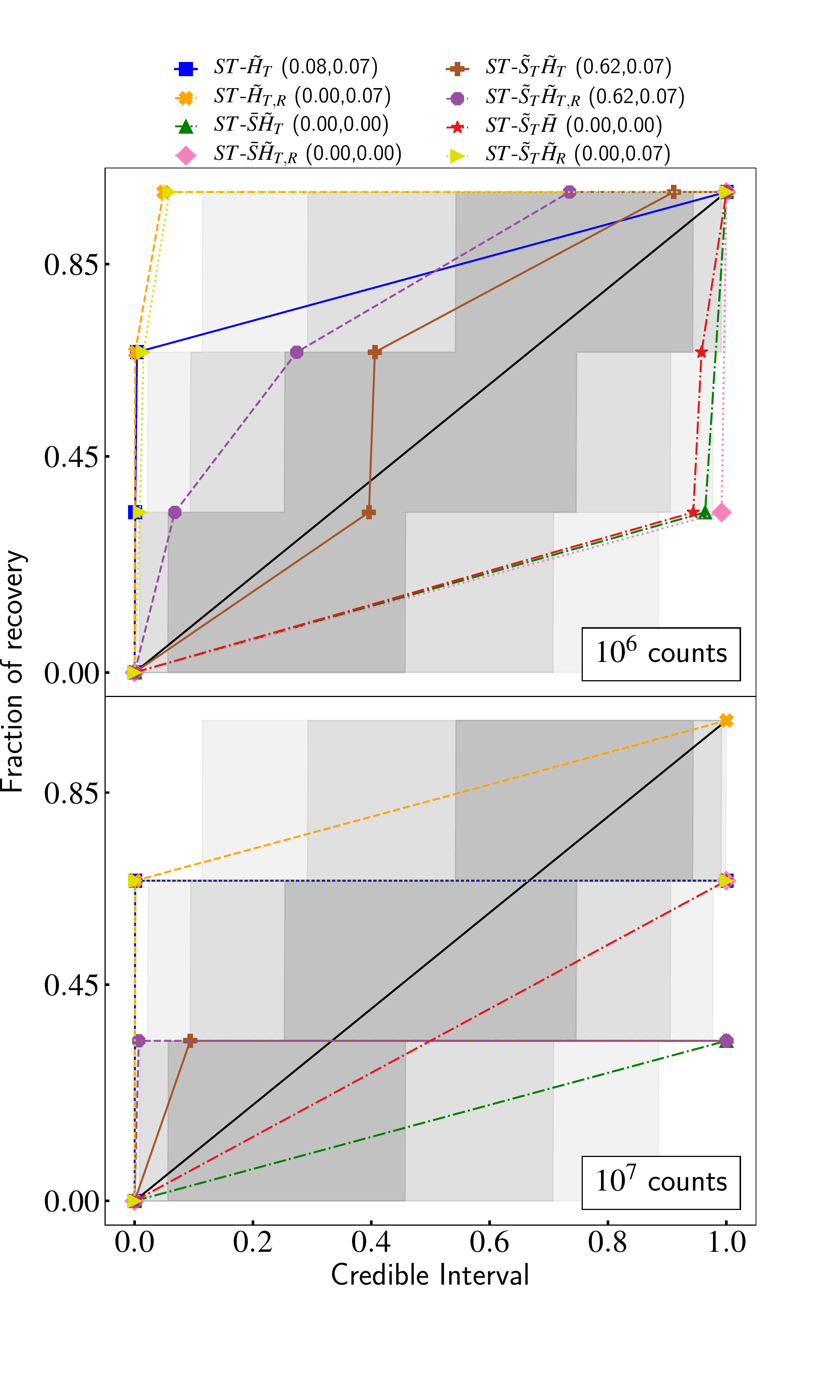}
    \caption{Cumulative fraction of compactness values recovered within a credible interval as a function of the credible intervals (from the lower tail of the posterior distribution) for each model using the \textit{variability data set}. The gray regions, from most opaque to least, depict the cumulative 1, 2, and 3$\sigma$ confidence intervals. The black solid line represents the expectation for unbiased parameter estimates. Each model and their corresponding \textit{p}-values are shown on top of the figure (left in the parenthesis: 10$^6$ counts, right in the parenthesis: 10$^7$ counts)}
    \label{fig:compactness_per_model}
    \end{figure}
    

    
\section{Discussion}\label{sec:discussion}

The major objective of this study was to examine parameter recovery for pulse profile modeling of TBOs: both ground truth recovery, and the uncertainties when the temporal variations that we expect from TBO sources are ignored. 
To achieve this objective, we used cutting-edge pulse profile models (incorporating the relevant general and special relativistic effects, bursting atmosphere physics, and interstellar absorption), coupled with a reasonable background model, and constructed realistic test cases based on one of the most promising sources for this type of analysis, J1814. 

We started by generating synthetic data that mimicked J1814 burst light curves and oscillation properties. In doing so, we developed eight phenomenological models that cover our current (highly uncertain) understanding of burning physics and in particular the physical mechanism that gives rise to the burst oscillations. 
Then, we created two data sets: a \textit{calibration data set} and a \textit{variability data set}. The \textit{calibration data set} contains no time variability and was primarily used to determine the optimal settings for our sampler and to assess the uncertainties that can be achieved on mass and radius estimates. The \textit{variability data set}, which contains bursts with time variability, was used to test the recovery of parameters when time variability (of the emitting region temperatures and sizes, depending on the model) is ignored during inference - an approach that could reduce computational cost of inference runs. To evaluate the quality of parameter recovery, we used PP plots, wrongness parameter (the difference between recovered and injected parameters in units of 68\% credible interval), posterior plots, and residuals. These tools allowed us to assess how well our sampler is able to recover the true values of the parameters in the presence of realistic levels of time variability.

\subsection{Phenomenological burst oscillation models}\label{sec:model_discussion}
We found that all of the phenomenological models considered are adequate to account for the main observed properties (an rms FA of the fundamental that is relatively constant throughout the burst and with an overall rms FA of 10\%) of the J1814 oscillation\footnote{Note that we did not attempt to match the harmonic content of the J1814 oscillations in our synthetic data generation, since our primary focus was on assessing the impact of variability.  The harmonic content of our synthetic data varied substantially.}. Thus, testing each of these models would be desirable when modeling actual J1814 data, depending on whether bursts are analyzed individually or merged  before analysis. 

Establishing a definitive link between these simple phenomenological models and the physical mechanisms responsible for oscillation
is something that has not yet been done, and certainly our phenomenological models do not fully encapsulate the intricacies of theoretical oscillation mechanisms.  Nevertheless, as detailed in Section \ref{sec:origins}, 
the simple models we proposed were motivated by physical models, and testing such phenomenological hot spot models via pulse profile modeling may eventually provide new insight into the burst oscillation mechanism.

\subsection{Mass and radius constraints}\label{sec:M-R}
Using our \textit{calibration data set}, we found that uncertainties on the mass and radius measurements can be reduced to 10\% and 15\% (width of the 68\% credible interval), respectively, assuming 10$^7$ counts from the star. This result is consistent with the findings of \citet{Lo:2013ava}, who predicted mass and radius constraints of 10\% when considering only 10$^6$ counts from a hot spot located within 10$^\circ$ of the equator. In contrast, a hot spot located within 20$^\circ$ of the pole provided no useful constraint in their analysis. Our analysis falls somewhere in between these two cases, as the location of the hot spot in our analysis was allowed to be anywhere in the northern hemisphere, as long as it resulted in a lightcurve and oscillations that mimicked J1814's  properties. The analysis performed on the \textit{calibration data set} also shows that mass has smaller fractional uncertainties than radius. This implies that the mass is typically more tightly constrained than the radius.

When considering the \textit{variability data set}, we determined that the parameter recovery for 10$^5$ counts was nominally acceptable. This is owing to the fact that the sampling uncertainties on the mass and radius estimations are larger than the model uncertainties at this level of data quality. The average uncertainty of the mass and radius estimations was determined to be 30\%, which is much higher than the approximately 10\% uncertainty required to differentiate between different EoS models \citep{Ozel:2009}. This implies two things. First, when performing inference on a burst of about 10$^5$ counts, temporal variability can be neglected and yet the inferred mass and radius will not show any measurable bias. Second, the mass and radius uncertainties from a 10$^5$ counts run are insufficient to discriminate between EOS. Collecting more counts is therefore required. However, when more counts are collected by merging several bursts ( yielding counts $\gtrapprox$ 10$^6$) and inference is performed  neglecting the time variability, the mass and radius are incorrectly estimated. This occurs because, given the quality of the data, the modeling uncertainties outweigh the sampling uncertainties. In this context, ignoring time variability, given a limited computation budget, is ineffective. To correctly estimate the main properties of a star, more advanced methods are required.

\subsection{Addressing inference bias associated with time variability}\label{sec:solution}

One of the main reasons for neglecting time variability is to reduce computational costs, which are known to be large, based on analysis of NICER RMP data using X-PSI \citep[up to several hundred thousand CPU hours for the most complex models, see for example][]{Riley:2019}.  One strategy to mitigate the bias introduced by time variable parameters is to have a good, physically-motivated, model of the temporal evolution of of the photospheric temperature. Parameterizing such a temperature profile in our simulation pipeline would speed up the analysis significantly. However getting such a temperature profile from nuclear theory is difficult in practice for a real source, given the complexity in modelling X-ray bursts.  To accurately model X-ray bursts, it is necessary to have comprehensive information about the thermal and compositional structure of the neutron star envelope, together with more than 1000 reaction rates involving over 300 different nuclides \citep[][and references therein]{Meisel:2018nmq}. 

In the absence of more accurate models, slicing the light curve into shorter segments where the variability can be neglected, and then jointly fitting the various segments, may reduce the systematic uncertainties. However, this approach is likely to come at considerable computational cost. Preliminary test runs reveal that while a run using the burst data of set 1 of our simplest model ST-$\tilde{H}_{T}$ takes approximately 13 000 core-hours, slicing  the same data into an arbitrary eight pieces, and doing a joint fit  allowing only the hot-spot temperature  to vary from slice to slice, results in the inference run taking approximately 150 000 core-hours. We plan to further explore this procedure in future work.

To get a sufficient number of counts to probe the EoS, it is necessary to combine many bursts, since for RXTE data  the number of photons in a single burst is not sufficient. Although we did not investigate the known burst to burst variability in our analysis, we anticipate that it will add to the temporal variability bias. Consequently, future X-ray missions: enhanced X-ray Timing and Polarimetry \citep[e-XTP;][]{Zhang:2019, Watts:2019_extp},
and the Spectroscopic Time-Resolving Observatory for Broadband Energy X-rays \citep[STROBE-X;][]{Ray:2019} missions, which will have larger effective areas covering the TBO energy range, may help us to avoid this additional complication by collecting sufficient number of counts in fewer bursts. However, accounting for the temporal variability when analyzing single bursts will still be required, even for these future instruments. With current modelling techniques, this would require allocation of a substantial computational budget.

\subsection{Some limitations of this study}\label{sec:caveat}
Despite the insights provided by our analysis, 
it is important to note that the derived uncertainties are conditional on the properties of J1814, which we used as a basis for our synthetic data sets. However, different TBO sources have distinct features, making it difficult to generalize our results.  In order to mitigate these uncertainties and increase the robustness of our findings, it may be necessary to conduct a more randomized analysis on a larger sample size. 

We are also not really able to reach a definitive conclusion about each individual model's performance using the \textit{variability data set}.   Some models that neglect time varibility may well perform better than others.  However with only three samples for each model (set by computational constraints), it is possible that the 
 poor performance of some models may be due to random fluctuation. 
 To draw more robust conclusions, it would be necessary to expand the number of parameter sets for each model.  Nevertheless, our results indicate that ignoring variability is risky. 

\subsection{Previous J1814 analysis via its TBO properties}\label{sec: prior_analysis}

The first analysis to constrain a the properties of a neutron star via pulse profile modelling of TBOs was conducted by  \citet{Bhattacharyya:2004pp} using J1814 bursts.
\citet{Bhattacharyya:2004pp} did not perform Bayesian inference in the way that we now do when pulse profile modelling (analysis methods have evolved substantially in the last few years in support of NICER), but they did use a relativistic ray-tracing model to fit a pulse profile obtained by combining - or `stacking' - data from multiple bursts. Doing this, 
the study derived constraints on J1814's compactness, hot spot, and geometrical properties. Several assumptions and simplifications were made during the analysis, including the following: (1) the entire analysis relied on two particular EoS (imposing a prior constraint on the mass-radius relation); (2) it was conditional on a hot spot with a fixed temperature of 2 keV; (3) the effect of the interstellar medium absorption was not included.  Most importantly, given the results of our new study, possible variability during and between the bursts that were stacked together was neglected. However, our study has found that ignoring variability during the bursts makes inference of the compactness problematic, especially when the number of stacked bursts yields a total counts of approximately 10$^6$ photons and above (as in the case of J1814).

\subsection{Future prospects}\label{sec: future}



In principle, pulse profile modeling of TBO sources should be able to deliver information not only on the mass and radius but also on the TBO spot properties, including their time evolution. As we have shown in this paper, extracting this information is currently challenging. Given the complexity in accurately modelling X-ray bursts and the uncertainty on the oscillations' origin, we proceeded in this analysis by testing different models relying on assumptions spanning a wide range of possibilities. Any theoretically motivated constraint would therefore increase the efficiency of future investigations. In particular, we would like to encourage the development of parameterized models of potential burst oscillation mechanisms, preferably coupled to models for the overall surface temperature evolution, in a form that can be used for pulse profile modeling so that we can test them more thoroughly. Alternatively, in the absence of a good physical model, slicing the burst into short time segments during which we can ignore the variability - while computationally intensive - may eventually help to constrain the wide range of possible models.
Further study will be needed to determine how well we can constrain models - what would it require, for example, to determine whether spot growth has a preferred direction rather than being symmetric?   We also need to give more consideration to how to model other components that might contribute to the data, such as pulsed emission from ongoing accretion-powered pulsations, or reprocessing of burst photons by the accretion flow.

Untangling these effects will be complicated, but there are also good prospects for using complementary constraints to help overcome these various issues.  Not all bursts from TBO sources have detectable oscillations, but phase-averaged spectral modelling of those bursts can still yield constraints on mass and radius \citep{Suleimanov:2011,Ozel:2013, Steiner:2013, Guillot:2014, Nattila:2016, Steiner:2018}, as can modelling of accretion disk phenomena like kilohertz QPOs \citep{Abolmasov:2020} or broadened iron lines \citep{Bhattacharyya:2011, Bhattacharyya:2017, Ludlam:2022}. Some TBO sources also have accretion-powered pulsations, for which pulse profile modeling can also be done \citep[e.g.][]{Salmi:2018}, yielding constraints on mass, radius and geometry.  Being able to perform precisely this kind of cross-check on high quality data is one of the drivers for future facilities such e-XTP and STROBE-X. X-ray polarimetry of such sources also offers a potential way to constrain geometries \citep{Viironen:2004,Watts:2019_extp,Loktev:2020, Salmi:2021}.  If we can put all of these pieces together, the prospects for reducing systematics and uncertainties are  highly promising.

\section*{Acknowledgements}

Y.K., T.S., A.L.W., S.V. and D.C. acknowledge support from ERC Consolidator Grant No.~865768 AEONS (PI Watts). The major part of the work was carried out on the HELIOS cluster including dedicated nodes funded via the abovementioned ERC CoG. Additionally, this work used the Dutch national e-infrastructure with the support of the SURF Cooperative using grant no. EINF-3731. S.F. contributed to this work as part of his Bachelor Project at the University of Amsterdam. We thank I. M. Romero-Shaw for discussion on PP-plots. We express our gratitude to Anna Bilous, Phil Uttley, Tess Jaffe, and Bryan K. Irby for providing their valuable assistance with regards to the response file. Z.M. was supported by the U.S. Department of Energy Office of Science under Grants No. DE-FG02-88ER40387 and DE- SC0019042, the U.S. National Nuclear Security Administration through Grant No. DE-NA0003909, and the U.S. National Science Foundation under Grant No. OISE-1927130 (International Research Network for Nuclear Astrophysics). V.S. acknowledges support by  Deutsche Forschungsgemeinschaft (DFG; grant WE 1312/59-1)

\section*{Data availability}
For this analysis, we used the publicly available simulation code X-PSI (\url{https://github.com/xpsi-group/xpsi.git}). The synthetic data and inference runs are both accessible on Zenodo (\url{https://doi.org/10.5281/zenodo.7665653}). 


\bibliographystyle{mnras}
\bibliography{bibliography}

\begin{thebibliography}{}
\makeatletter
\relax
\def\mn@urlcharsother{\let\do\@makeother \do\$\do\&\do\#\do\^\do\_\do\%\do\~}
\def\mn@doi{\begingroup\mn@urlcharsother \@ifnextchar [ {\mn@doi@}
  {\mn@doi@[]}}
\def\mn@doi@[#1]#2{\def\@tempa{#1}\ifx\@tempa\@empty \href
  {http://dx.doi.org/#2} {doi:#2}\else \href {http://dx.doi.org/#2} {#1}\fi
  \endgroup}
\def\mn@eprint#1#2{\mn@eprint@#1:#2::\@nil}
\def\mn@eprint@arXiv#1{\href {http://arxiv.org/abs/#1} {{\tt arXiv:#1}}}
\def\mn@eprint@dblp#1{\href {http://dblp.uni-trier.de/rec/bibtex/#1.xml}
  {dblp:#1}}
\def\mn@eprint@#1:#2:#3:#4\@nil{\def\@tempa {#1}\def\@tempb {#2}\def\@tempc
  {#3}\ifx \@tempc \@empty \let \@tempc \@tempb \let \@tempb \@tempa \fi \ifx
  \@tempb \@empty \def\@tempb {arXiv}\fi \@ifundefined
  {mn@eprint@\@tempb}{\@tempb:\@tempc}{\expandafter \expandafter \csname
  mn@eprint@\@tempb\endcsname \expandafter{\@tempc}}}

\bibitem[\protect\citeauthoryear{{Abolmasov}, {N{\"a}ttil{\"a}}  \&
  {Poutanen}}{{Abolmasov} et~al.}{2020}]{Abolmasov:2020}
{Abolmasov} P.,  {N{\"a}ttil{\"a}} J.,   {Poutanen} J.,  2020, \mn@doi [\aap]
  {10.1051/0004-6361/201936958}, \href
  {https://ui.adsabs.harvard.edu/abs/2020A&A...638A.142A} {638, A142}

\bibitem[\protect\citeauthoryear{{Altamirano} et~al.,}{{Altamirano}
  et~al.}{2010}]{Altamirano:2010}
{Altamirano} D.,  et~al., 2010, The Astronomer's Telegram, \href
  {https://ui.adsabs.harvard.edu/abs/2010ATel.2932....1A} {2932, 1}

\bibitem[\protect\citeauthoryear{{Baub{\"o}ck}, {Psaltis}, {{\"O}zel}  \&
  {Johannsen}}{{Baub{\"o}ck} et~al.}{2012}]{Baubock:2012}
{Baub{\"o}ck} M.,  {Psaltis} D.,  {{\"O}zel} F.,   {Johannsen} T.,  2012,
  \mn@doi [\apj] {10.1088/0004-637X/753/2/175}, \href
  {https://ui.adsabs.harvard.edu/abs/2012ApJ...753..175B} {753, 175}

\bibitem[\protect\citeauthoryear{Baym, Hatsuda, Kojo, Powell, Song  \&
  Takatsuka}{Baym et~al.}{2018}]{Baym:2017}
Baym G.,  Hatsuda T.,  Kojo T.,  Powell P.~D.,  Song Y.,   Takatsuka T.,  2018,
  \mn@doi [Rept. Prog. Phys.] {10.1088/1361-6633/aaae14}, 81, 056902

\bibitem[\protect\citeauthoryear{{Belian}, {Conner}  \& {Evans}}{{Belian}
  et~al.}{1976}]{Belian:1976}
{Belian} R.~D.,  {Conner} J.~P.,   {Evans} W.~D.,  1976, \mn@doi [\apjl]
  {10.1086/182151}, \href
  {https://ui.adsabs.harvard.edu/abs/1976ApJ...206L.135B} {206, L135}

\bibitem[\protect\citeauthoryear{{Beloborodov}}{{Beloborodov}}{2002}]{Beloborodov:2002}
{Beloborodov} A.~M.,  2002, \mn@doi [\apjl] {10.1086/339511}, \href
  {https://ui.adsabs.harvard.edu/abs/2002ApJ...566L..85B} {566, L85}

\bibitem[\protect\citeauthoryear{{Berry} et~al.,}{{Berry}
  et~al.}{2015}]{Berry:2015}
{Berry} C. P.~L.,  et~al., 2015, \mn@doi [\apj] {10.1088/0004-637X/804/2/114},
  \href {https://ui.adsabs.harvard.edu/abs/2015ApJ...804..114B} {804, 114}

\bibitem[\protect\citeauthoryear{{Bhattacharyya}}{{Bhattacharyya}}{2011}]{Bhattacharyya:2011}
{Bhattacharyya} S.,  2011, \mn@doi [\mnras] {10.1111/j.1365-2966.2011.18936.x},
  \href {https://ui.adsabs.harvard.edu/abs/2011MNRAS.415.3247B} {415, 3247}

\bibitem[\protect\citeauthoryear{{Bhattacharyya}}{{Bhattacharyya}}{2017}]{Bhattacharyya:2017}
{Bhattacharyya} S.,  2017, \mn@doi [Journal of Astrophysics and Astronomy]
  {10.1007/s12036-017-9459-4}, \href
  {https://ui.adsabs.harvard.edu/abs/2017JApA...38...38B} {38, 38}

\bibitem[\protect\citeauthoryear{{Bhattacharyya}}{{Bhattacharyya}}{2022}]{Bhattacharyya:2022}
{Bhattacharyya} S.,  2022, in {Bhattacharyya} S.,  {Papitto} A.,
  {Bhattacharya} D.,  eds,  Astrophysics and Space Science Library Vol. 465,
  Astrophysics and Space Science Library. pp 125--155 (\mn@eprint {arXiv}
  {2103.11258}), \mn@doi{10.1007/978-3-030-85198-9_5}

\bibitem[\protect\citeauthoryear{{Bhattacharyya} \&
  {Strohmayer}}{{Bhattacharyya} \& {Strohmayer}}{2006}]{Bhattacharyya:2006b}
{Bhattacharyya} S.,  {Strohmayer} T.~E.,  2006, \mn@doi [\apjl]
  {10.1086/500199}, \href
  {https://ui.adsabs.harvard.edu/abs/2006ApJ...636L.121B} {636, L121}

\bibitem[\protect\citeauthoryear{Bhattacharyya, Strohmayer, Miller  \&
  Markwardt}{Bhattacharyya et~al.}{2005}]{Bhattacharyya:2004pp}
Bhattacharyya S.,  Strohmayer T.~E.,  Miller M.~C.,   Markwardt C.~B.,  2005,
  \mn@doi [\apj] {10.1086/426383}, 619, 483

\bibitem[\protect\citeauthoryear{{Bildsten}}{{Bildsten}}{1998}]{Bildsten:1998}
{Bildsten} L.,  1998, in {Buccheri} R.,  {van Paradijs} J.,   {Alpar} A.,  eds,
   NATO Advanced Study Institute (ASI) Series C Vol. 515, The Many Faces of
  Neutron Stars.. p.~419 (\mn@eprint {arXiv} {astro-ph/9709094})

\bibitem[\protect\citeauthoryear{Bilous \& Watts}{Bilous \&
  Watts}{2019}]{Bilous:2018fxm}
Bilous A.~V.,  Watts A.~L.,  2019, \mn@doi [\apjs] {10.3847/1538-4365/ab2fe1},
  245, 19

\bibitem[\protect\citeauthoryear{{Bilous} et~al.,}{{Bilous}
  et~al.}{2019}]{Bilous:2019}
{Bilous} A.~V.,  et~al., 2019, \mn@doi [\apjl] {10.3847/2041-8213/ab53e7},
  \href {https://ui.adsabs.harvard.edu/abs/2019ApJ...887L..23B} {887, L23}

\bibitem[\protect\citeauthoryear{{Biwer}, {Capano}, {De}, {Cabero}, {Brown},
  {Nitz}  \& {Raymond}}{{Biwer} et~al.}{2019}]{Biwer:2019}
{Biwer} C.~M.,  {Capano} C.~D.,  {De} S.,  {Cabero} M.,  {Brown} D.~A.,  {Nitz}
  A.~H.,   {Raymond} V.,  2019, \mn@doi [\pasp] {10.1088/1538-3873/aaef0b},
  \href {https://ui.adsabs.harvard.edu/abs/2019PASP..131b4503B} {131, 024503}

\bibitem[\protect\citeauthoryear{{Bogdanov} et~al.,}{{Bogdanov}
  et~al.}{2019}]{Bogdanov:2019}
{Bogdanov} S.,  et~al., 2019, \mn@doi [\apjl] {10.3847/2041-8213/ab5968}, \href
  {https://ui.adsabs.harvard.edu/abs/2019ApJ...887L..26B} {887, L26}

\bibitem[\protect\citeauthoryear{{Bogdanov} et~al.,}{{Bogdanov}
  et~al.}{2021}]{Bogdanov:2021}
{Bogdanov} S.,  et~al., 2021, \mn@doi [\apjl] {10.3847/2041-8213/abfb79}, \href
  {https://ui.adsabs.harvard.edu/abs/2021ApJ...914L..15B} {914, L15}

\bibitem[\protect\citeauthoryear{{Braje}, {Romani}  \& {Rauch}}{{Braje}
  et~al.}{2000}]{Braje:2000}
{Braje} T.~M.,  {Romani} R.~W.,   {Rauch} K.~P.,  2000, \mn@doi [\apj]
  {10.1086/308448}, \href
  {https://ui.adsabs.harvard.edu/abs/2000ApJ...531..447B} {531, 447}

\bibitem[\protect\citeauthoryear{{Cadeau}, {Morsink}, {Leahy}  \&
  {Campbell}}{{Cadeau} et~al.}{2007}]{Cadeau:2007}
{Cadeau} C.,  {Morsink} S.~M.,  {Leahy} D.,   {Campbell} S.~S.,  2007, \mn@doi
  [\apj] {10.1086/509103}, \href
  {https://ui.adsabs.harvard.edu/abs/2007ApJ...654..458C} {654, 458}

\bibitem[\protect\citeauthoryear{{Cameron}}{{Cameron}}{2011}]{Cameron:2011}
{Cameron} E.,  2011, \mn@doi [\pasa] {10.1071/AS10046}, \href
  {https://ui.adsabs.harvard.edu/abs/2011PASA...28..128C} {28, 128}

\bibitem[\protect\citeauthoryear{{Campana}}{{Campana}}{2022}]{Campana:2022}
{Campana} R.,  2022, arXiv e-prints, \href
  {https://ui.adsabs.harvard.edu/abs/2022arXiv220509359C} {p. arXiv:2205.09359}

\bibitem[\protect\citeauthoryear{{Cavecchi} \& {Spitkovsky}}{{Cavecchi} \&
  {Spitkovsky}}{2019}]{Cavecchi:2019}
{Cavecchi} Y.,  {Spitkovsky} A.,  2019, \mn@doi [\apj]
  {10.3847/1538-4357/ab3650}, \href
  {https://ui.adsabs.harvard.edu/abs/2019ApJ...882..142C} {882, 142}

\bibitem[\protect\citeauthoryear{{Cavecchi} et~al.,}{{Cavecchi}
  et~al.}{2011}]{Cavecchi:2011}
{Cavecchi} Y.,  et~al., 2011, \mn@doi [\apjl] {10.1088/2041-8205/740/1/L8},
  \href {https://ui.adsabs.harvard.edu/abs/2011ApJ...740L...8C} {740, L8}

\bibitem[\protect\citeauthoryear{{Chambers} \& {Watts}}{{Chambers} \&
  {Watts}}{2020}]{Chambers:2020}
{Chambers} F.~R.~N.,  {Watts} A.~L.,  2020, \mn@doi [\mnras]
  {10.1093/mnras/stz3449}, \href
  {https://ui.adsabs.harvard.edu/abs/2020MNRAS.491.6032C} {491, 6032}

\bibitem[\protect\citeauthoryear{{Chambers}, {Watts}, {Keek}, {Cavecchi}  \&
  {Garcia}}{{Chambers} et~al.}{2019}]{Chambers:2019}
{Chambers} F. R.~N.,  {Watts} A.~L.,  {Keek} L.,  {Cavecchi} Y.,   {Garcia} F.,
   2019, \mn@doi [\apj] {10.3847/1538-4357/aaf501}, \href
  {https://ui.adsabs.harvard.edu/abs/2019ApJ...871...61C} {871, 61}

\bibitem[\protect\citeauthoryear{{Chandrasekhar}}{{Chandrasekhar}}{1960}]{Cha60}
{Chandrasekhar} S.,  1960, {Radiative transfer}.
Dover, New York

\bibitem[\protect\citeauthoryear{{Chandrasekhar} \& {Breen}}{{Chandrasekhar} \&
  {Breen}}{1947}]{Cha47}
{Chandrasekhar} S.,  {Breen} F.~H.,  1947, \mn@doi [\apj] {10.1086/144918},
  \href {http://adsabs.harvard.edu/abs/1947ApJ...105..435C} {105, 435}

\bibitem[\protect\citeauthoryear{{Chen} \& {Shaham}}{{Chen} \&
  {Shaham}}{1989}]{Chen:1989}
{Chen} K.,  {Shaham} J.,  1989, \mn@doi [\apj] {10.1086/167295}, \href
  {https://ui.adsabs.harvard.edu/abs/1989ApJ...339..279C} {339, 279}

\bibitem[\protect\citeauthoryear{{Chen}, {Yuan}  \& {Vasilopoulos}}{{Chen}
  et~al.}{2020}]{Chen:2020}
{Chen} A.~Y.,  {Yuan} Y.,   {Vasilopoulos} G.,  2020, \mn@doi [\apjl]
  {10.3847/2041-8213/ab85c5}, \href
  {https://ui.adsabs.harvard.edu/abs/2020ApJ...893L..38C} {893, L38}

\bibitem[\protect\citeauthoryear{Cook, Gelman  \& Rubin}{Cook
  et~al.}{2006}]{cook_2006}
Cook S.~R.,  Gelman A.,   Rubin D.~B.,  2006, \mn@doi [Journal of Computational
  and Graphical Statistics] {10.1198/106186006X136976}, 15, 675

\bibitem[\protect\citeauthoryear{{Cornish}, {Littenberg}, {B{\'e}csy},
  {Chatziioannou}, {Clark}, {Ghonge}  \& {Millhouse}}{{Cornish}
  et~al.}{2021}]{Cornish:2021}
{Cornish} N.~J.,  {Littenberg} T.~B.,  {B{\'e}csy} B.,  {Chatziioannou} K.,
  {Clark} J.~A.,  {Ghonge} S.,   {Millhouse} M.,  2021, \mn@doi [\prd]
  {10.1103/PhysRevD.103.044006}, \href
  {https://ui.adsabs.harvard.edu/abs/2021PhRvD.103d4006C} {103, 044006}

\bibitem[\protect\citeauthoryear{{Ding}, {Deller}  \& {Miller-Jones}}{{Ding}
  et~al.}{2021}]{Ding:2021}
{Ding} H.,  {Deller} A.~T.,   {Miller-Jones} J. C.~A.,  2021, \mn@doi [\pasa]
  {10.1017/pasa.2021.37}, \href
  {https://ui.adsabs.harvard.edu/abs/2021PASA...38...48D} {38, e048}

\bibitem[\protect\citeauthoryear{{Ebisuzaki}}{{Ebisuzaki}}{1987}]{Ebisuzaki:1987}
{Ebisuzaki} T.,  1987, \pasj, \href
  {https://ui.adsabs.harvard.edu/abs/1987PASJ...39..287E} {39, 287}

\bibitem[\protect\citeauthoryear{{Feroz} \& {Hobson}}{{Feroz} \&
  {Hobson}}{2008}]{MultiNest_2008}
{Feroz} F.,  {Hobson} M.~P.,  2008, \mn@doi [\mnras]
  {10.1111/j.1365-2966.2007.12353.x}, \href
  {https://ui.adsabs.harvard.edu/abs/2008MNRAS.384..449F} {384, 449}

\bibitem[\protect\citeauthoryear{{Feroz}, {Hobson}  \& {Bridges}}{{Feroz}
  et~al.}{2009}]{MultiNest_2009}
{Feroz} F.,  {Hobson} M.~P.,   {Bridges} M.,  2009, \mn@doi [\mnras]
  {10.1111/j.1365-2966.2009.14548.x}, \href
  {http://adsabs.harvard.edu/abs/2009MNRAS.398.1601F} {398, 1601}

\bibitem[\protect\citeauthoryear{{Feroz}, {Hobson}, {Cameron}  \&
  {Pettitt}}{{Feroz} et~al.}{2019}]{MultiNest_2019}
{Feroz} F.,  {Hobson} M.~P.,  {Cameron} E.,   {Pettitt} A.~N.,  2019, \mn@doi
  [The Open Journal of Astrophysics] {10.21105/astro.1306.2144}, \href
  {https://ui.adsabs.harvard.edu/abs/2019OJAp....2E..10F} {2, 10}

\bibitem[\protect\citeauthoryear{{Gaia Collaboration} et~al.,}{{Gaia
  Collaboration} et~al.}{2016}]{Gaia_Collaboration}
{Gaia Collaboration} et~al., 2016, \mn@doi [\aap]
  {10.1051/0004-6361/201629272}, \href
  {https://ui.adsabs.harvard.edu/abs/2016A&A...595A...1G} {595, A1}

\bibitem[\protect\citeauthoryear{{Gair} \& {Moore}}{{Gair} \&
  {Moore}}{2015}]{Gair:2015}
{Gair} J.~R.,  {Moore} C.~J.,  2015, \mn@doi [\prd]
  {10.1103/PhysRevD.91.124062}, \href
  {https://ui.adsabs.harvard.edu/abs/2015PhRvD..91l4062G} {91, 124062}

\bibitem[\protect\citeauthoryear{Galloway \& Keek}{Galloway \&
  Keek}{2020}]{Galloway:2017}
Galloway D.~K.,  Keek L.,  2020, \mn@doi [Astrophys. Space Sci. Libr.]
  {10.1007/978-3-662-62110-3_5}, 461, 209

\bibitem[\protect\citeauthoryear{Galloway et~al.,}{Galloway
  et~al.}{2020}]{Galloway:2020}
Galloway D.~K.,  et~al., 2020, \mn@doi [\apjs] {10.3847/1538-4365/ab9f2e}, 249,
  32

\bibitem[\protect\citeauthoryear{Garcia, Chambers  \& Watts}{Garcia
  et~al.}{2019}]{Garcia:2018}
Garcia F.,  Chambers F. R.~N.,   Watts A.~L.,  2019, \mn@doi [Phys. Rev.
  Fluids.] {10.1103/PhysRevFluids.4.074802}, 4, 074802

\bibitem[\protect\citeauthoryear{{Gendreau} et~al.,}{{Gendreau}
  et~al.}{2016}]{NICER}
{Gendreau} K.~C.,  et~al., 2016, in {den Herder} J.-W.~A.,  {Takahashi} T.,
  {Bautz} M.,  eds,  Society of Photo-Optical Instrumentation Engineers (SPIE)
  Conference Series Vol. 9905, Space Telescopes and Instrumentation 2016:
  Ultraviolet to Gamma Ray. p. 99051H, \mn@doi{10.1117/12.2231304}

\bibitem[\protect\citeauthoryear{Gough}{Gough}{2009}]{Gough:2009}
Gough B.,  2009, GNU Scientific Library Reference Manual - Third Edition, 3rd
  edn.
Network Theory Ltd.

\bibitem[\protect\citeauthoryear{{Gralla}, {Lupsasca}  \& {Philippov}}{{Gralla}
  et~al.}{2017}]{Gralla:2017}
{Gralla} S.~E.,  {Lupsasca} A.,   {Philippov} A.,  2017, \mn@doi [\apj]
  {10.3847/1538-4357/aa978d}, \href
  {https://ui.adsabs.harvard.edu/abs/2017ApJ...851..137G} {851, 137}

\bibitem[\protect\citeauthoryear{{Grindlay}, {Gursky}, {Schnopper},
  {Parsignault}, {Heise}, {Brinkman}  \& {Schrijver}}{{Grindlay}
  et~al.}{1976}]{Grindlay:1976}
{Grindlay} J.,  {Gursky} H.,  {Schnopper} H.,  {Parsignault} D.~R.,  {Heise}
  J.,  {Brinkman} A.~C.,   {Schrijver} J.,  1976, \mn@doi [\apjl]
  {10.1086/182105}, \href
  {https://ui.adsabs.harvard.edu/abs/1976ApJ...205L.127G} {205, L127}

\bibitem[\protect\citeauthoryear{{Guillot} \& {Rutledge}}{{Guillot} \&
  {Rutledge}}{2014}]{Guillot:2014}
{Guillot} S.,  {Rutledge} R.~E.,  2014, \mn@doi [\apjl]
  {10.1088/2041-8205/796/1/L3}, \href
  {https://ui.adsabs.harvard.edu/abs/2014ApJ...796L...3G} {796, L3}

\bibitem[\protect\citeauthoryear{{Harding} \& {Muslimov}}{{Harding} \&
  {Muslimov}}{2001}]{Harding:2001}
{Harding} A.~K.,  {Muslimov} A.~G.,  2001, \mn@doi [\apj] {10.1086/321589},
  \href {https://ui.adsabs.harvard.edu/abs/2001ApJ...556..987H} {556, 987}

\bibitem[\protect\citeauthoryear{{Harding} \& {Muslimov}}{{Harding} \&
  {Muslimov}}{2002}]{Harding:2002}
{Harding} A.~K.,  {Muslimov} A.~G.,  2002, \mn@doi [\apj] {10.1086/338985},
  \href {https://ui.adsabs.harvard.edu/abs/2002ApJ...568..862H} {568, 862}

\bibitem[\protect\citeauthoryear{Hebeler}{Hebeler}{2021}]{Hebeler:2020}
Hebeler K.,  2021, \mn@doi [Phys. Rept.] {10.1016/j.physrep.2020.08.009}, 890,
  1

\bibitem[\protect\citeauthoryear{{Heyl}}{{Heyl}}{2004}]{Heyl:2004}
{Heyl} J.~S.,  2004, \mn@doi [\apj] {10.1086/379966}, \href
  {https://ui.adsabs.harvard.edu/abs/2004ApJ...600..939H} {600, 939}

\bibitem[\protect\citeauthoryear{{Jahoda}, {Swank}, {Giles}, {Stark},
  {Strohmayer}, {Zhang}  \& {Morgan}}{{Jahoda} et~al.}{1996}]{Jahoda:1996}
{Jahoda} K.,  {Swank} J.~H.,  {Giles} A.~B.,  {Stark} M.~J.,  {Strohmayer} T.,
  {Zhang} W.,   {Morgan} E.~H.,  1996, in {Siegmund} O.~H.,  {Gummin} M.~A.,
  eds,  Society of Photo-Optical Instrumentation Engineers (SPIE) Conference
  Series Vol. 2808, EUV, X-Ray, and Gamma-Ray Instrumentation for Astronomy
  VII. pp 59--70, \mn@doi{10.1117/12.256034}

\bibitem[\protect\citeauthoryear{{Kalapotharakos}, {Wadiasingh}, {Harding}  \&
  {Kazanas}}{{Kalapotharakos} et~al.}{2021}]{Kalapotharakos:2021}
{Kalapotharakos} C.,  {Wadiasingh} Z.,  {Harding} A.~K.,   {Kazanas} D.,  2021,
  \mn@doi [\apj] {10.3847/1538-4357/abcec0}, \href
  {https://ui.adsabs.harvard.edu/abs/2021ApJ...907...63K} {907, 63}

\bibitem[\protect\citeauthoryear{Kini et~al.,}{Kini
  et~al.}{2023}]{zenodo_paper}
Kini Y.,  et~al., 2023, {Pulse Profile Modeling of Thermonuclear Burst
  Oscillations I: The Effect of Neglecting Variability},
  \mn@doi{10.5281/zenodo.7665653}, \url
  {https://doi.org/10.5281/zenodo.7665653}

\bibitem[\protect\citeauthoryear{Krauss et~al.,}{Krauss
  et~al.}{2005}]{Krauss:2005sj}
Krauss M.~I.,  et~al., 2005, \mn@doi [\apj] {10.1086/430595}, 627, 910

\bibitem[\protect\citeauthoryear{{Lapidus}, {Syunyaev}  \&
  {Titarchuk}}{{Lapidus} et~al.}{1986}]{Lapidus:1986}
{Lapidus} I.~I.,  {Syunyaev} R.~A.,   {Titarchuk} L.~G.,  1986, Soviet
  Astronomy Letters, \href
  {https://ui.adsabs.harvard.edu/abs/1986SvAL...12..383L} {12, 383}

\bibitem[\protect\citeauthoryear{Lattimer}{Lattimer}{2012}]{Lattimer:2012}
Lattimer J.~M.,  2012, \mn@doi [Ann. Rev. Nucl. Part. Sci.]
  {10.1146/annurev-nucl-102711-095018}, 62, 485

\bibitem[\protect\citeauthoryear{{Lee}}{{Lee}}{2004}]{Lee:2004}
{Lee} U.,  2004, \mn@doi [\apj] {10.1086/380122}, \href
  {https://ui.adsabs.harvard.edu/abs/2004ApJ...600..914L} {600, 914}

\bibitem[\protect\citeauthoryear{Lo, Coleman~Miller, Bhattacharyya  \& Lamb}{Lo
  et~al.}{2013}]{Lo:2013ava}
Lo K.~H.,  Coleman~Miller M.,  Bhattacharyya S.,   Lamb F.~K.,  2013, \mn@doi
  [\apj] {10.1088/0004-637X/776/1/19}, 776, 19

\bibitem[\protect\citeauthoryear{{Loktev}, {Salmi}, {N{\"a}ttil{\"a}}  \&
  {Poutanen}}{{Loktev} et~al.}{2020}]{Loktev:2020}
{Loktev} V.,  {Salmi} T.,  {N{\"a}ttil{\"a}} J.,   {Poutanen} J.,  2020,
  \mn@doi [\aap] {10.1051/0004-6361/202039134}, \href
  {https://ui.adsabs.harvard.edu/abs/2020A&A...643A..84L} {643, A84}

\bibitem[\protect\citeauthoryear{{London}, {Howard}  \& {Taam}}{{London}
  et~al.}{1984}]{London:1984}
{London} R.~A.,  {Howard} W.~M.,   {Taam} R.~E.,  1984, \mn@doi [\apjl]
  {10.1086/184390}, \href
  {https://ui.adsabs.harvard.edu/abs/1984ApJ...287L..27L} {287, L27}

\bibitem[\protect\citeauthoryear{{London}, {Taam}  \& {Howard}}{{London}
  et~al.}{1986}]{London:1986}
{London} R.~A.,  {Taam} R.~E.,   {Howard} W.~M.,  1986, \mn@doi [\apj]
  {10.1086/164330}, \href
  {https://ui.adsabs.harvard.edu/abs/1986ApJ...306..170L} {306, 170}

\bibitem[\protect\citeauthoryear{{Ludlam} et~al.,}{{Ludlam}
  et~al.}{2022}]{Ludlam:2022}
{Ludlam} R.~M.,  et~al., 2022, \mn@doi [\apj] {10.3847/1538-4357/ac5028}, \href
  {https://ui.adsabs.harvard.edu/abs/2022ApJ...927..112L} {927, 112}

\bibitem[\protect\citeauthoryear{{Madej}}{{Madej}}{1991}]{Madej:1991}
{Madej} J.,  1991, \mn@doi [\apj] {10.1086/170264}, \href
  {https://ui.adsabs.harvard.edu/abs/1991ApJ...376..161M} {376, 161}

\bibitem[\protect\citeauthoryear{{Madej}, {Joss}  \&
  {R{\'o}{\.z}a{\'n}ska}}{{Madej} et~al.}{2004}]{Madej:2004}
{Madej} J.,  {Joss} P.~C.,   {R{\'o}{\.z}a{\'n}ska} A.,  2004, \mn@doi [\apj]
  {10.1086/379761}, \href
  {https://ui.adsabs.harvard.edu/abs/2004ApJ...602..904M} {602, 904}

\bibitem[\protect\citeauthoryear{{Mahmoodifar} \& {Strohmayer}}{{Mahmoodifar}
  \& {Strohmayer}}{2016}]{Mahmoodifar:2016}
{Mahmoodifar} S.,  {Strohmayer} T.,  2016, \mn@doi [\apj]
  {10.3847/0004-637X/818/1/93}, \href
  {https://ui.adsabs.harvard.edu/abs/2016ApJ...818...93M} {818, 93}

\bibitem[\protect\citeauthoryear{Meisel}{Meisel}{2018}]{Meisel:2018rsy}
Meisel Z.,  2018, \mn@doi [\apj] {10.3847/1538-4357/aac3d3}, 860, 147

\bibitem[\protect\citeauthoryear{Meisel, Merz  \& Medvid}{Meisel
  et~al.}{2019}]{Meisel:2018nmq}
Meisel Z.,  Merz G.,   Medvid S.,  2019, \mn@doi [\apj]
  {10.3847/1538-4357/aafede}, 872, 84

\bibitem[\protect\citeauthoryear{{Miller} \& {Lamb}}{{Miller} \&
  {Lamb}}{1998}]{Miller:1998}
{Miller} M.~C.,  {Lamb} F.~K.,  1998, \mn@doi [\apjl] {10.1086/311335}, \href
  {https://ui.adsabs.harvard.edu/abs/1998ApJ...499L..37M} {499, L37}

\bibitem[\protect\citeauthoryear{{Miller} \& {Lamb}}{{Miller} \&
  {Lamb}}{2015}]{Miller:2015}
{Miller} M.~C.,  {Lamb} F.~K.,  2015, \mn@doi [\apj]
  {10.1088/0004-637X/808/1/31}, \href
  {https://ui.adsabs.harvard.edu/abs/2015ApJ...808...31M} {808, 31}

\bibitem[\protect\citeauthoryear{Miller et~al.}{Miller
  et~al.}{2019}]{Miller:2019}
Miller M.~C.,  et~al., 2019, \mn@doi [\apjl] {10.3847/2041-8213/ab50c5}, 887,
  L24

\bibitem[\protect\citeauthoryear{Miller et~al.}{Miller
  et~al.}{2021}]{Miller:2021}
Miller M.~C.,  et~al., 2021, \mn@doi [\apjl] {10.3847/2041-8213/ac089b}, 918,
  L28

\bibitem[\protect\citeauthoryear{{Moran}, {Mingarelli}, {Bedell}  \&
  {Good}}{{Moran} et~al.}{2022}]{Moran:2022}
{Moran} A.,  {Mingarelli} C. M.~F.,  {Bedell} M.,   {Good} D.,  2022, arXiv
  e-prints, \href {https://ui.adsabs.harvard.edu/abs/2022arXiv221010816M} {p.
  arXiv:2210.10816}

\bibitem[\protect\citeauthoryear{{Morsink}, {Leahy}, {Cadeau}  \&
  {Braga}}{{Morsink} et~al.}{2007}]{Morsink:2007}
{Morsink} S.~M.,  {Leahy} D.~A.,  {Cadeau} C.,   {Braga} J.,  2007, \mn@doi
  [\apj] {10.1086/518648}, \href
  {https://ui.adsabs.harvard.edu/abs/2007ApJ...663.1244M} {663, 1244}

\bibitem[\protect\citeauthoryear{{Mozzon}, {Ashton}, {Nuttall}  \&
  {Williamson}}{{Mozzon} et~al.}{2022}]{Mozzon:2022}
{Mozzon} S.,  {Ashton} G.,  {Nuttall} L.~K.,   {Williamson} A.~R.,  2022,
  \mn@doi [\prd] {10.1103/PhysRevD.106.043504}, \href
  {https://ui.adsabs.harvard.edu/abs/2022PhRvD.106d3504M} {106, 043504}

\bibitem[\protect\citeauthoryear{{Muno}, {{\"O}zel}  \& {Chakrabarty}}{{Muno}
  et~al.}{2002}]{Muno:2002}
{Muno} M.~P.,  {{\"O}zel} F.,   {Chakrabarty} D.,  2002, \mn@doi [\apj]
  {10.1086/344152}, \href
  {https://ui.adsabs.harvard.edu/abs/2002ApJ...581..550M} {581, 550}

\bibitem[\protect\citeauthoryear{{N{\"a}ttil{\"a}} \&
  {Pihajoki}}{{N{\"a}ttil{\"a}} \& {Pihajoki}}{2018}]{Nattila:2018}
{N{\"a}ttil{\"a}} J.,  {Pihajoki} P.,  2018, \mn@doi [\aap]
  {10.1051/0004-6361/201630261}, \href
  {https://ui.adsabs.harvard.edu/abs/2018A&A...615A..50N} {615, A50}

\bibitem[\protect\citeauthoryear{{N{\"a}ttil{\"a}}, {Steiner}, {Kajava},
  {Suleimanov}  \& {Poutanen}}{{N{\"a}ttil{\"a}} et~al.}{2016}]{Nattila:2016}
{N{\"a}ttil{\"a}} J.,  {Steiner} A.~W.,  {Kajava} J.~J.~E.,  {Suleimanov}
  V.~F.,   {Poutanen} J.,  2016, \mn@doi [\aap] {10.1051/0004-6361/201527416},
  \href {https://ui.adsabs.harvard.edu/abs/2016A&A...591A..25N} {591, A25}

\bibitem[\protect\citeauthoryear{Oertel, Hempel, Kl\"ahn  \& Typel}{Oertel
  et~al.}{2017}]{Oertel:2016}
Oertel M.,  Hempel M.,  Kl\"ahn T.,   Typel S.,  2017, \mn@doi [Rev. Mod.
  Phys.] {10.1103/RevModPhys.89.015007}, 89, 015007

\bibitem[\protect\citeauthoryear{{{\"O}zel}}{{{\"O}zel}}{2013}]{Ozel:2013}
{{\"O}zel} F.,  2013, \mn@doi [Reports on Progress in Physics]
  {10.1088/0034-4885/76/1/016901}, \href
  {https://ui.adsabs.harvard.edu/abs/2013RPPh...76a6901O} {76, 016901}

\bibitem[\protect\citeauthoryear{{{\"O}zel} \& {Psaltis}}{{{\"O}zel} \&
  {Psaltis}}{2009}]{Ozel:2009}
{{\"O}zel} F.,  {Psaltis} D.,  2009, \mn@doi [\prd]
  {10.1103/PhysRevD.80.103003}, \href
  {https://ui.adsabs.harvard.edu/abs/2009PhRvD..80j3003O} {80, 103003}

\bibitem[\protect\citeauthoryear{{Page}}{{Page}}{1995}]{Page:1995}
{Page} D.,  1995, \mn@doi [\apj] {10.1086/175439}, \href
  {https://ui.adsabs.harvard.edu/abs/1995ApJ...442..273P} {442, 273}

\bibitem[\protect\citeauthoryear{{Pankow}, {Brady}, {Ochsner}  \&
  {O'Shaughnessy}}{{Pankow} et~al.}{2015}]{Pankow:2015}
{Pankow} C.,  {Brady} P.,  {Ochsner} E.,   {O'Shaughnessy} R.,  2015, \mn@doi
  [\prd] {10.1103/PhysRevD.92.023002}, \href
  {https://ui.adsabs.harvard.edu/abs/2015PhRvD..92b3002P} {92, 023002}

\bibitem[\protect\citeauthoryear{{Paxton}, {Bildsten}, {Dotter}, {Herwig},
  {Lesaffre}  \& {Timmes}}{{Paxton} et~al.}{2011}]{Paxton:2011}
{Paxton} B.,  {Bildsten} L.,  {Dotter} A.,  {Herwig} F.,  {Lesaffre} P.,
  {Timmes} F.,  2011, \mn@doi [\apjs] {10.1088/0067-0049/192/1/3}, \href
  {https://ui.adsabs.harvard.edu/abs/2011ApJS..192....3P} {192, 3}

\bibitem[\protect\citeauthoryear{{Pechenick}, {Ftaclas}  \&
  {Cohen}}{{Pechenick} et~al.}{1983}]{Pechenick:1983}
{Pechenick} K.~R.,  {Ftaclas} C.,   {Cohen} J.~M.,  1983, \mn@doi [\apj]
  {10.1086/161498}, \href
  {https://ui.adsabs.harvard.edu/abs/1983ApJ...274..846P} {274, 846}

\bibitem[\protect\citeauthoryear{{Philippov} \& {Spitkovsky}}{{Philippov} \&
  {Spitkovsky}}{2018}]{Philippov:2018}
{Philippov} A.~A.,  {Spitkovsky} A.,  2018, \mn@doi [\apj]
  {10.3847/1538-4357/aaabbc}, \href
  {https://ui.adsabs.harvard.edu/abs/2018ApJ...855...94P} {855, 94}

\bibitem[\protect\citeauthoryear{{Piro} \& {Bildsten}}{{Piro} \&
  {Bildsten}}{2005}]{Piro:2005}
{Piro} A.~L.,  {Bildsten} L.,  2005, \mn@doi [\apj] {10.1086/430777}, \href
  {https://ui.adsabs.harvard.edu/abs/2005ApJ...629..438P} {629, 438}

\bibitem[\protect\citeauthoryear{{Poutanen} \& {Beloborodov}}{{Poutanen} \&
  {Beloborodov}}{2006}]{Poutanen:2006}
{Poutanen} J.,  {Beloborodov} A.~M.,  2006, \mn@doi [\mnras]
  {10.1111/j.1365-2966.2006.11088.x}, \href
  {https://ui.adsabs.harvard.edu/abs/2006MNRAS.373..836P} {373, 836}

\bibitem[\protect\citeauthoryear{{Poutanen} \& {Gierli{\'n}ski}}{{Poutanen} \&
  {Gierli{\'n}ski}}{2003}]{Poutanen:2003}
{Poutanen} J.,  {Gierli{\'n}ski} M.,  2003, \mn@doi [\mnras]
  {10.1046/j.1365-8711.2003.06773.x}, \href
  {https://ui.adsabs.harvard.edu/abs/2003MNRAS.343.1301P} {343, 1301}

\bibitem[\protect\citeauthoryear{Psaltis, \"Ozel  \& Chakrabarty}{Psaltis
  et~al.}{2014}]{Psaltis:2013fha}
Psaltis D.,  \"Ozel F.,   Chakrabarty D.,  2014, \mn@doi [\apj]
  {10.1088/0004-637X/787/2/136}, 787, 136

\bibitem[\protect\citeauthoryear{{Ray} et~al.,}{{Ray} et~al.}{2019}]{Ray:2019}
{Ray} P.~S.,  et~al., 2019, \mn@doi [arXiv e-prints]
  {10.48550/arXiv.1903.03035}, \href
  {https://ui.adsabs.harvard.edu/abs/2019arXiv190303035R} {p. arXiv:1903.03035}

\bibitem[\protect\citeauthoryear{{Riley}, {Raaijmakers}  \& {Watts}}{{Riley}
  et~al.}{2018}]{Riley:2018}
{Riley} T.~E.,  {Raaijmakers} G.,   {Watts} A.~L.,  2018, \mn@doi [\mnras]
  {10.1093/mnras/sty1051}, \href
  {https://ui.adsabs.harvard.edu/abs/2018MNRAS.478.1093R} {478, 1093}

\bibitem[\protect\citeauthoryear{Riley et~al.}{Riley et~al.}{2019}]{Riley:2019}
Riley T.~E.,  et~al., 2019, \mn@doi [\apjl] {10.3847/2041-8213/ab481c}, 887,
  L21

\bibitem[\protect\citeauthoryear{Riley et~al.}{Riley et~al.}{2021}]{Riley:2021}
Riley T.~E.,  et~al., 2021, \mn@doi [\apjl] {10.3847/2041-8213/ac0a81}, 918,
  L27

\bibitem[\protect\citeauthoryear{Riley et~al.,}{Riley
  et~al.}{2023a}]{xpsi_v122_zenodo}
Riley T.~E.,  et~al., 2023a, {X-PSI: A Python package for neutron star X-ray
  pulse simulation and inference}, \mn@doi{10.5281/zenodo.7632629}, \url
  {https://doi.org/10.5281/zenodo.7632629}

\bibitem[\protect\citeauthoryear{Riley et~al.,}{Riley
  et~al.}{2023b}]{Riley2023}
Riley T.~E.,  et~al., 2023b, \mn@doi [Journal of Open Source Software]
  {10.21105/joss.04977}, 8, 4977

\bibitem[\protect\citeauthoryear{{Romero-Shaw} et~al.,}{{Romero-Shaw}
  et~al.}{2020}]{Romero-Shaw:2020}
{Romero-Shaw} I.~M.,  et~al., 2020, \mn@doi [\mnras] {10.1093/mnras/staa2850},
  \href {https://ui.adsabs.harvard.edu/abs/2020MNRAS.499.3295R} {499, 3295}

\bibitem[\protect\citeauthoryear{{Salmi}, {N{\"a}ttil{\"a}}  \&
  {Poutanen}}{{Salmi} et~al.}{2018}]{Salmi:2018}
{Salmi} T.,  {N{\"a}ttil{\"a}} J.,   {Poutanen} J.,  2018, \mn@doi [\aap]
  {10.1051/0004-6361/201833348}, \href
  {https://ui.adsabs.harvard.edu/abs/2018A&A...618A.161S} {618, A161}

\bibitem[\protect\citeauthoryear{{Salmi}, {Loktev}, {Korsman}, {Baldini},
  {Tsygankov}  \& {Poutanen}}{{Salmi} et~al.}{2021}]{Salmi:2021}
{Salmi} T.,  {Loktev} V.,  {Korsman} K.,  {Baldini} L.,  {Tsygankov} S.~S.,
  {Poutanen} J.,  2021, \mn@doi [\aap] {10.1051/0004-6361/202039470}, \href
  {https://ui.adsabs.harvard.edu/abs/2021A&A...646A..23S} {646, A23}

\bibitem[\protect\citeauthoryear{{Salmi} et~al.,}{{Salmi}
  et~al.}{2022}]{Salmi:2022}
{Salmi} T.,  et~al., 2022, \mn@doi [\apj] {10.3847/1538-4357/ac983d}, \href
  {https://ui.adsabs.harvard.edu/abs/2022ApJ...941..150S} {941, 150}

\bibitem[\protect\citeauthoryear{{Schatz} et~al.,}{{Schatz}
  et~al.}{2001}]{Schatz:2001}
{Schatz} H.,  et~al., 2001, \mn@doi [\prl] {10.1103/PhysRevLett.86.3471}, \href
  {https://ui.adsabs.harvard.edu/abs/2001PhRvL..86.3471S} {86, 3471}

\bibitem[\protect\citeauthoryear{{Sidery} et~al.,}{{Sidery}
  et~al.}{2014}]{Sidery:2014}
{Sidery} T.,  et~al., 2014, \mn@doi [\prd] {10.1103/PhysRevD.89.084060}, \href
  {https://ui.adsabs.harvard.edu/abs/2014PhRvD..89h4060S} {89, 084060}

\bibitem[\protect\citeauthoryear{{Sobolev}}{{Sobolev}}{1949}]{sob49}
{Sobolev} V.~V.,  1949, Uch. Zap. Leningrad Univ., 16

\bibitem[\protect\citeauthoryear{{Sobolev}}{{Sobolev}}{1963}]{Sob63}
{Sobolev} V.~V.,  1963, {A treatise on radiative transfer}.
Van Nostrand, Princeton

\bibitem[\protect\citeauthoryear{{Spitkovsky}, {Levin}  \&
  {Ushomirsky}}{{Spitkovsky} et~al.}{2002}]{Spitkovsky:2002}
{Spitkovsky} A.,  {Levin} Y.,   {Ushomirsky} G.,  2002, \mn@doi [\apj]
  {10.1086/338040}, \href
  {https://ui.adsabs.harvard.edu/abs/2002ApJ...566.1018S} {566, 1018}

\bibitem[\protect\citeauthoryear{{Steiner}, {Lattimer}  \& {Brown}}{{Steiner}
  et~al.}{2013}]{Steiner:2013}
{Steiner} A.~W.,  {Lattimer} J.~M.,   {Brown} E.~F.,  2013, \mn@doi [\apjl]
  {10.1088/2041-8205/765/1/L5}, \href
  {https://ui.adsabs.harvard.edu/abs/2013ApJ...765L...5S} {765, L5}

\bibitem[\protect\citeauthoryear{{Steiner}, {Heinke}, {Bogdanov}, {Li}, {Ho},
  {Bahramian}  \& {Han}}{{Steiner} et~al.}{2018}]{Steiner:2018}
{Steiner} A.~W.,  {Heinke} C.~O.,  {Bogdanov} S.,  {Li} C.~K.,  {Ho} W.~C.~G.,
  {Bahramian} A.,   {Han} S.,  2018, \mn@doi [\mnras] {10.1093/mnras/sty215},
  \href {https://ui.adsabs.harvard.edu/abs/2018MNRAS.476..421S} {476, 421}

\bibitem[\protect\citeauthoryear{{Stevens}, {Fiege}, {Leahy}  \&
  {Morsink}}{{Stevens} et~al.}{2016}]{Stevens:2016}
{Stevens} A.~L.,  {Fiege} J.~D.,  {Leahy} D.~A.,   {Morsink} S.~M.,  2016,
  \mn@doi [\apj] {10.3847/1538-4357/833/2/244}, \href
  {https://ui.adsabs.harvard.edu/abs/2016ApJ...833..244S} {833, 244}

\bibitem[\protect\citeauthoryear{{Strohmayer} \& {Markwardt}}{{Strohmayer} \&
  {Markwardt}}{2010}]{Strohmayer:2010}
{Strohmayer} T.~E.,  {Markwardt} C.~B.,  2010, The Astronomer's Telegram, \href
  {https://ui.adsabs.harvard.edu/abs/2010ATel.2929....1S} {2929, 1}

\bibitem[\protect\citeauthoryear{{Strohmayer}, {Zhang}, {Swank}, {Smale},
  {Titarchuk}, {Day}  \& {Lee}}{{Strohmayer} et~al.}{1996}]{Strohmayer:1996}
{Strohmayer} T.~E.,  {Zhang} W.,  {Swank} J.~H.,  {Smale} A.,  {Titarchuk} L.,
  {Day} C.,   {Lee} U.,  1996, \mn@doi [\apjl] {10.1086/310261}, \href
  {https://ui.adsabs.harvard.edu/abs/1996ApJ...469L...9S} {469, L9}

\bibitem[\protect\citeauthoryear{{Strohmayer}, {Zhang}  \&
  {Swank}}{{Strohmayer} et~al.}{1997}]{Strohmayer:1997}
{Strohmayer} T.~E.,  {Zhang} W.,   {Swank} J.~H.,  1997, \mn@doi [\apjl]
  {10.1086/310880}, \href
  {https://ui.adsabs.harvard.edu/abs/1997ApJ...487L..77S} {487, L77}

\bibitem[\protect\citeauthoryear{Strohmayer, Markwardt, Swank  \& in~'t
  Zand}{Strohmayer et~al.}{2003}]{Strohmayer:2003}
Strohmayer T.~E.,  Markwardt C.~B.,  Swank J.~H.,   in~'t Zand J.,  2003,
  \mn@doi [\apjl] {10.1086/379158}, 596, L67

\bibitem[\protect\citeauthoryear{{Suleimanov}, {Poutanen}  \&
  {Werner}}{{Suleimanov} et~al.}{2011a}]{Valery2011}
{Suleimanov} V.,  {Poutanen} J.,   {Werner} K.,  2011a, \mn@doi [\aap]
  {10.1051/0004-6361/201015845}, \href
  {https://ui.adsabs.harvard.edu/abs/2011A&A...527A.139S} {527, A139}

\bibitem[\protect\citeauthoryear{{Suleimanov}, {Poutanen}, {Revnivtsev}  \&
  {Werner}}{{Suleimanov} et~al.}{2011b}]{Suleimanov:2011}
{Suleimanov} V.,  {Poutanen} J.,  {Revnivtsev} M.,   {Werner} K.,  2011b,
  \mn@doi [\apj] {10.1088/0004-637X/742/2/122}, \href
  {https://ui.adsabs.harvard.edu/abs/2011ApJ...742..122S} {742, 122}

\bibitem[\protect\citeauthoryear{{Suleimanov}, {Poutanen}  \&
  {Werner}}{{Suleimanov} et~al.}{2012}]{Valery2012}
{Suleimanov} V.,  {Poutanen} J.,   {Werner} K.,  2012, \mn@doi [\aap]
  {10.1051/0004-6361/201219480}, \href
  {https://ui.adsabs.harvard.edu/abs/2012A&A...545A.120S} {545, A120}

\bibitem[\protect\citeauthoryear{Suleimanov, Poutanen, N\"attil\"a, Kajava,
  Revnivtsev  \& Werner}{Suleimanov et~al.}{2017}]{Valery2017}
Suleimanov V.~F.,  Poutanen J.,  N\"attil\"a J.,  Kajava J. J.~E.,  Revnivtsev
  M.~G.,   Werner K.,  2017, \mn@doi [\mnras] {10.1093/mnras/stw3132}, 466, 906

\bibitem[\protect\citeauthoryear{{Suleimanov}, {Poutanen}  \&
  {Werner}}{{Suleimanov} et~al.}{2018}]{Valery2018}
{Suleimanov} V.~F.,  {Poutanen} J.,   {Werner} K.,  2018, \mn@doi [\aap]
  {10.1051/0004-6361/201833581}, \href
  {http://adsabs.harvard.edu/abs/2018A%26A...619A.114S} {619, A114}

\bibitem[\protect\citeauthoryear{{Suleimanov}, {Poutanen}  \&
  {Werner}}{{Suleimanov} et~al.}{2020}]{Valery2020}
{Suleimanov} V.~F.,  {Poutanen} J.,   {Werner} K.,  2020, \mn@doi [\aap]
  {10.1051/0004-6361/202037502}, \href
  {https://ui.adsabs.harvard.edu/abs/2020arXiv200509759S} {639, A33}

\bibitem[\protect\citeauthoryear{Tolos \& Fabbietti}{Tolos \&
  Fabbietti}{2020}]{Tolos:2020}
Tolos L.,  Fabbietti L.,  2020, \mn@doi [Prog. Part. Nucl. Phys.]
  {10.1016/j.ppnp.2020.103770}, 112, 103770

\bibitem[\protect\citeauthoryear{{Veitch} et~al.,}{{Veitch}
  et~al.}{2015}]{Veitch:2015}
{Veitch} J.,  et~al., 2015, \mn@doi [\prd] {10.1103/PhysRevD.91.042003}, \href
  {https://ui.adsabs.harvard.edu/abs/2015PhRvD..91d2003V} {91, 042003}

\bibitem[\protect\citeauthoryear{{Viironen} \& {Poutanen}}{{Viironen} \&
  {Poutanen}}{2004}]{Viironen:2004}
{Viironen} K.,  {Poutanen} J.,  2004, \mn@doi [\aap]
  {10.1051/0004-6361:20041084}, \href
  {https://ui.adsabs.harvard.edu/abs/2004A&A...426..985V} {426, 985}

\bibitem[\protect\citeauthoryear{{Watts}}{{Watts}}{2012}]{Watts:2012}
{Watts} A.~L.,  2012, \mn@doi [\araa] {10.1146/annurev-astro-040312-132617},
  \href {https://ui.adsabs.harvard.edu/abs/2012ARA&A..50..609W} {50, 609}

\bibitem[\protect\citeauthoryear{{Watts}}{{Watts}}{2019a}]{Watts:2019}
{Watts} A.~L.,  2019a, in Xiamen-CUSTIPEN Workshop on the Equation of State of
  Dense Neutron-Rich Matter in the Era of Gravitational Wave Astronomy. p.
  020008 (\mn@eprint {arXiv} {1904.07012}), \mn@doi{10.1063/1.5117798}

\bibitem[\protect\citeauthoryear{Watts}{Watts}{2019b}]{Watts:2019lbs}
Watts A.~L.,  2019b, \mn@doi [AIP Conf. Proc.] {10.1063/1.5117798}, 2127,
  020008

\bibitem[\protect\citeauthoryear{Watts \& Strohmayer}{Watts \&
  Strohmayer}{2006}]{Watts:2006}
Watts A.~L.,  Strohmayer T.~E.,  2006, \mn@doi [\mnras]
  {10.1111/j.1365-2966.2006.11072.x}, 373, 769

\bibitem[\protect\citeauthoryear{{Watts}, {Strohmayer}  \& {Markwardt}}{{Watts}
  et~al.}{2005}]{Watts:2005}
{Watts} A.~L.,  {Strohmayer} T.~E.,   {Markwardt} C.~B.,  2005, \mn@doi [\apj]
  {10.1086/496953}, \href
  {https://ui.adsabs.harvard.edu/abs/2005ApJ...634..547W} {634, 547}

\bibitem[\protect\citeauthoryear{{Watts} et~al.,}{{Watts}
  et~al.}{2016}]{Watts:2016}
{Watts} A.~L.,  et~al., 2016, \mn@doi [Reviews of Modern Physics]
  {10.1103/RevModPhys.88.021001}, \href
  {https://ui.adsabs.harvard.edu/abs/2016RvMP...88b1001W} {88, 021001}

\bibitem[\protect\citeauthoryear{{Watts} et~al.,}{{Watts}
  et~al.}{2019}]{Watts:2019_extp}
{Watts} A.~L.,  et~al., 2019, \mn@doi [Science China Physics, Mechanics, and
  Astronomy] {10.1007/s11433-017-9188-4}, \href
  {https://ui.adsabs.harvard.edu/abs/2019SCPMA..6229503W} {62, 29503}

\bibitem[\protect\citeauthoryear{{Weinberg}, {Miller}  \& {Lamb}}{{Weinberg}
  et~al.}{2001}]{Weinberg:2001}
{Weinberg} N.,  {Miller} M.~C.,   {Lamb} D.~Q.,  2001, \mn@doi [\apj]
  {10.1086/318279}, \href
  {https://ui.adsabs.harvard.edu/abs/2001ApJ...546.1098W} {546, 1098}

\bibitem[\protect\citeauthoryear{{Wilms}, {Allen}  \& {McCray}}{{Wilms}
  et~al.}{2000}]{Wilms:2000}
{Wilms} J.,  {Allen} A.,   {McCray} R.,  2000, \mn@doi [\apj] {10.1086/317016},
  \href {https://ui.adsabs.harvard.edu/abs/2000ApJ...542..914W} {542, 914}

\bibitem[\protect\citeauthoryear{Worpel, Galloway  \& Price}{Worpel
  et~al.}{2013}]{Worpel:2013fka}
Worpel H.,  Galloway D.~K.,   Price D.~J.,  2013, \mn@doi [\apj]
  {10.1088/0004-637X/772/2/94}, 772, 94

\bibitem[\protect\citeauthoryear{Yang \& Piekarewicz}{Yang \&
  Piekarewicz}{2020}]{Yang:2019}
Yang J.,  Piekarewicz J.,  2020, \mn@doi [Ann. Rev. Nucl. Part. Sci.]
  {10.1146/annurev-nucl-101918-023608}, 70, 21

\bibitem[\protect\citeauthoryear{{You}, {Ashton}, {Zhu}, {Thrane}  \&
  {Zhu}}{{You} et~al.}{2022}]{You:2022}
{You} Z.-Q.,  {Ashton} G.,  {Zhu} X.-J.,  {Thrane} E.,   {Zhu} Z.-H.,  2022,
  \mn@doi [\mnras] {10.1093/mnras/stab2977}, \href
  {https://ui.adsabs.harvard.edu/abs/2022MNRAS.509.3957Y} {509, 3957}

\bibitem[\protect\citeauthoryear{{Zhang} et~al.,}{{Zhang}
  et~al.}{2019}]{Zhang:2019}
{Zhang} S.,  et~al., 2019, \mn@doi [Science China Physics, Mechanics, and
  Astronomy] {10.1007/s11433-018-9309-2}, \href
  {https://ui.adsabs.harvard.edu/abs/2019SCPMA..6229502Z} {62, 29502}

\bibitem[\protect\citeauthoryear{{in't Zand}}{{in't Zand}}{2017}]{Zand:2017}
{in't Zand} J.,  2017, in {Serino} M.,  {Shidatsu} M.,  {Iwakiri} W.,
  {Mihara} T.,  eds, 7 years of MAXI: monitoring X-ray Transients. p.~121
  (\mn@eprint {arXiv} {1702.04899})

\makeatother
\end{thebibliography}



\appendix


\section{Appendix }\label{sec:appendix}
\clearpage
\onecolumn
     
    Table \ref{tab:params_var} and \ref{tab:params_const} show the values used to generate synthetic data for each model for a given set of parameters. These values were obtained by following the methodology outlined in Section \ref{sec:method}. The models shown in Table \ref{tab:params_const} correspond to their variability counterparts, based on which synthetic data were generated using the method outlined in Section 1.

\begin{figure*}
    \hspace*{-1cm} 
    \centering
    \includegraphics[width=1.1\columnwidth]{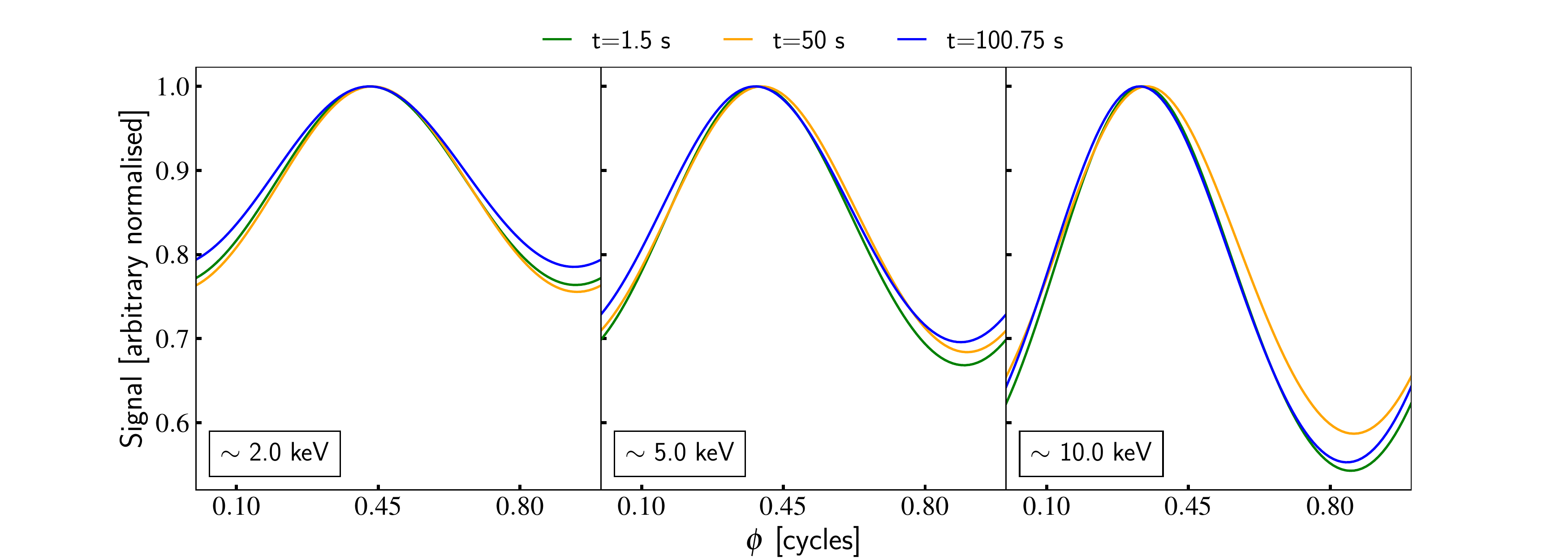}
    \caption{Three snapshots showing the evolution of the pulse profile corresponding to the light curve in Figure \ref{fig:burst}.}
    \label{fig:pulse}
    \end{figure*}

    In Figure \ref{fig:pulse}, we show the evolving pulse profile of parameters set 2 of model ST-$\bar{S}\tilde{H}_{T,R}$ (same as in Figure \ref{fig:burst}) at different energies. It is not surprising that the pulses remain relatively stable throughout the burst, given that having a consistent rms FA was one of the criteria used to select the parameter vector. However, the slight variations observed in the pulse at different times can be attributed to changes in the angular radius and temperature of the hotspot. For this burst, the inference was performed using 2-D `energy channel vs rotational phase'  the pulse profile between t = 1.5 s and t = 100.75 s, which corresponds the chopped data (see Section \ref{subsec:synthetic_data}).

\begin{figure}
    \centering
    \includegraphics[width=0.49\columnwidth]{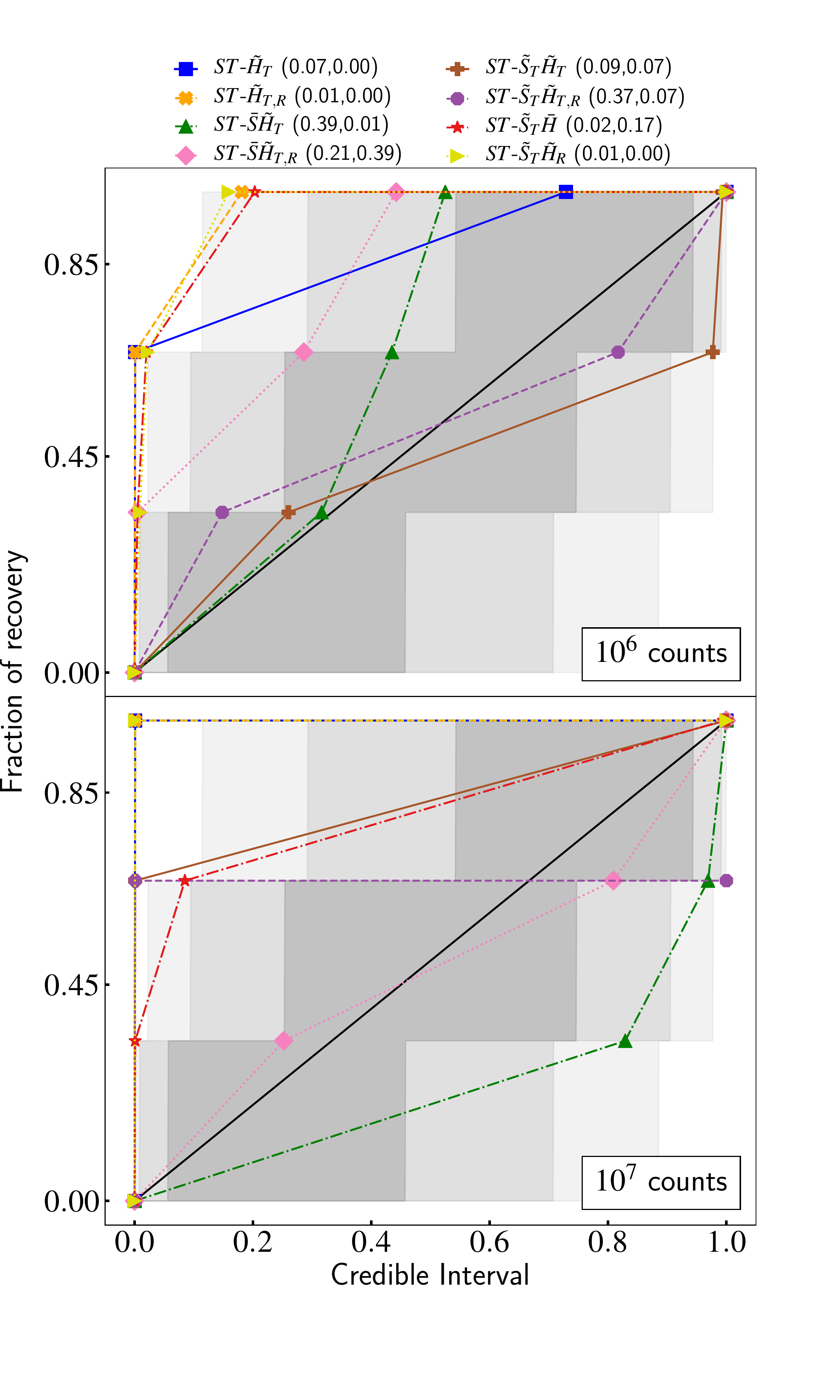}
    \includegraphics[width=0.49\columnwidth]{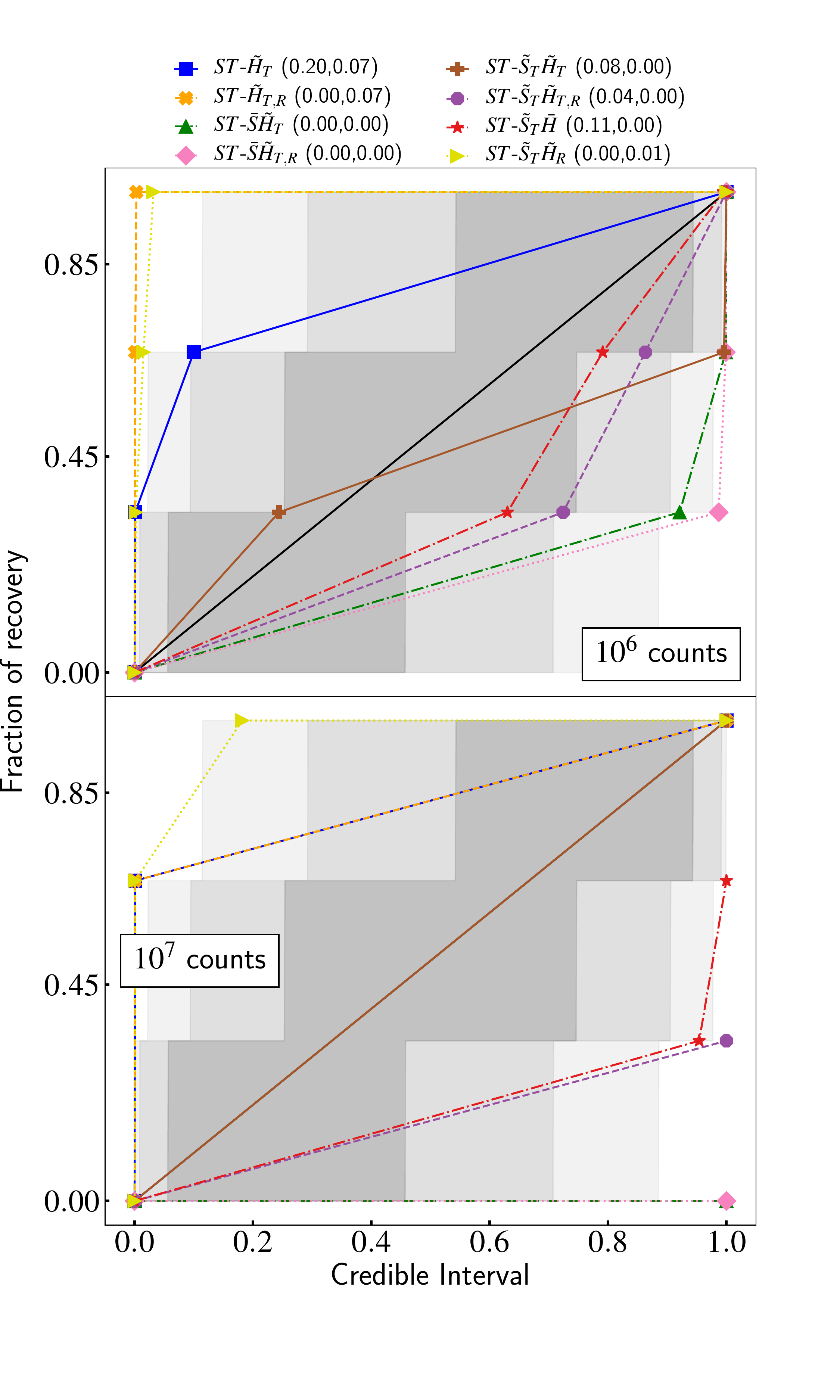}
    \caption{Radius (left) and mass (right) cumulative fraction of parameter values recovered within a credible interval as a function of the credible intervals for each model using the \textit{variability data set}. The gray regions, from most opaque to least, depict the cumulative 1, 2, and 3$\sigma$ confidence intervals. The black solid line represents the expectation for unbiased parameter estimates. Each model and their corresponding \textit{p}-values are shown on top of the figure.}
    \label{fig:radius_per_model}
    \end{figure}

    The cumulative fraction of parameter values recovered within a credible interval as a function of the credible intervals for each model using the \textit{variability data set} is shown in Figure \ref{fig:radius_per_model}. The radius plots are shown in the left panel, while the mass plots are shown in the right panel. The top panel represents 10$^6$ counts for each parameter, while the bottom panel represents 10$^7$ counts.

\begingroup
  \small \renewcommand{\arraystretch}{1.} \setlength{\tabcolsep}{2.pt}
  \begin{longtable}[c]{cccc|ccc|ccc|ccc|ccc|ccc|ccc|ccc|}

 \cline{2-25} \cline{2-25}
& \multicolumn{24}{|c|}{Models} \\
  \cline{2-25}

& \multicolumn{3}{{|c|}}{ST-$\tilde{H}_{T}$} & \multicolumn{3}{c|}{ST-$\tilde{H}_{T,R}$} & \multicolumn{3}{c|}{ST-$\bar{S}\tilde{H}_{T}$} & \multicolumn{3}{c|}{ST-$\bar{S}\tilde{H}_{T,R}$} & \multicolumn{3}{c|}{ST-$\tilde{S}_{T}\tilde{H}_{T}$} & \multicolumn{3}{c|}{ST-$\tilde{S}_{T}\tilde{H}_{T,R}$} & \multicolumn{3}{c|}{ST-$\tilde{S}_{T}\bar{H}$} & \multicolumn{3}{c|}{ST-$\tilde{S}_{T}\tilde{H}_{R}$} \\

\hline
\multicolumn{1}{|l|}{Parameter}       &s1  &s2   &s3   &s1   &s2   &s3   &s1   &s2    &s3   &s1   &s2   &s3   &s1  &s2  &s3  &s1  &s2  &s3  & s1 &s2  & s3 &s1  &s2  &s3 \\
\hline
\multicolumn{1}{|l|}{M ($M_{\odot}$)}               &1.5  & 2.1 & 1.3 &2.0  &1.2  & 1.7 &1.4  &1.4   &1.8  &1.2  &2.3  &1.7  &1.3 &2.3 &1.9 &1.6 &2.3 &1.2 &1.45&2.0 &1.3 &1.6 &1.4 &1.2\\ 
\multicolumn{1}{|l|}{R (km)}               &13.0 &12.3 & 12.2&14.5 &12.5 & 14.0&10.0 &10.0  &14.5 &9.0  &10.5 &12.0 &6.0 &10.0&8.26&8.0 &10.0&6.0 &12.5&12.0&12.2&10  &7.0 &8.75\\
\multicolumn{1}{|l|}{D (kpc)}               &7.5  & 3.5 & 1.0 &5.8  &6.1  &6.2  &5.5  &5.7   &8    &5.3  &3.5  &7.0  &5.0 &8.0 &6.5 &6.5 &7.0 &5.2 &11.1&8.6 &12.0&10  &7.0 &9.0\\
\multicolumn{1}{|l|}{\textit{i}}      &30   & 13  & 12  &25   &25   &30   &45   &43    &35   &30   &45   &35   &45  &50  &60  &60  &70  &60  &35  &30  &85  &55  &50  &55    \\
\multicolumn{1}{|l|}{$\phi_\mathrm{spot}$ (cycles)}   &0.3  & 0.6 & 0.9 &0.3  &0.6  &0.9  &0.3  &0.6   &0.9  &0.8  &0.6  &0.9  &0.3 &0.6 &0.9 &0.3 &0.6 &0.9 &0.3 &0.6 &0.9 &0.0 &0.6 &0.9   \\
\multicolumn{1}{|l|}{$\theta_\mathrm{spot}$ (radian)} &20   & 47  & 42  &25   &20   &22   &20   &20    &20   &25   &25  &20   &65  &55  &60  &55  &70  &60  &40  &60  &40  &55  &50  &55    \\
\multicolumn{1}{|l|}{$\zeta_\mathrm{spot}$ (radian)}  &0.75 & 0.5 & 0.1 &\var &\var &\var &1.0  &1.0   &0.75 &\var &\var &\var &0.75&1.0 &0.75&\var&\var&\var&0.75&0.95&0.75&\var&\var&\var \\
\multicolumn{1}{|l|}{$\log[T_\mathrm{spot}(\mathrm{K})/1\mathrm{K}]$}      &\var &\var &\var &\var &\var &\var &\var &\var  &\var &\var &\var &\var &\var&\var&\var&\var&\var&\var&7.4 &7.5 &7.5 &7.4 &7.45&7.4     \\
\multicolumn{1}{|l|}{$\log[T_\mathrm{star}(\mathrm{K})/1\mathrm{K}]$ }      &\none&\none&\none&\none&\none&\none&6.83 &6.83  &6.83 &6.83 &6.83 &6.83 &\var&\var&\var&\var&\var&\var&\var&\var&\var&\var&\var&\var  \\
\multicolumn{1}{|l|}{$N_H$ ($10^{20} \mathrm{cm}^{-2}$)}          & 1   & 3   &6    &7    &2    &2    &2    &5     &3    &2    &8    &8    &1   &9   &5   &6   &6   &6   &1   &5   &5   &1   &1   &1\\

\hline
\caption{Parameters used to produce the for \textit{variability data set};\var: time dependent parameter, \none : parameter not used for this case. sX denotes parameter set X  of the corresponding model, X $\in \{1,2,3\}$}.
 \label{tab:params_var}

\end{longtable}
\endgroup

\begingroup
 \small \renewcommand{\arraystretch}{1.} \setlength{\tabcolsep}{1.7pt}
  \begin{longtable}[c]{cccc|ccc|ccc|ccc|ccc|ccc|ccc|ccc|}


 \cline{2-25} \cline{2-25}
& \multicolumn{24}{|c|}{Models} \\
  \cline{2-25}

& \multicolumn{3}{{|c|}}{ST-$\tilde{H}_{T}$} & \multicolumn{3}{c|}{ST-$\tilde{H}_{T,R}$} & \multicolumn{3}{c|}{ST-$\bar{S}\tilde{H}_{T}$} & \multicolumn{3}{c|}{ST-$\bar{S}\tilde{H}_{T,R}$} & \multicolumn{3}{c|}{ST-$\tilde{S}_{T}\tilde{H}_{T}$}  & \multicolumn{3}{c|}{ST-$\tilde{S}_{T}\tilde{H}_{T,R}$} & \multicolumn{3}{c|}{ST-$\tilde{S}_{T}\bar{H}$} & \multicolumn{3}{c|}{ST-$\tilde{S}_{T}\tilde{H}_{R}$} \\

\hline
\multicolumn{1}{|l|}{Parameter}       &s1  &s2   &s3   &s1   &s2   &s3   &s1   &s2    &s3   &s1   &s2   &s3   &s1  &s2  &s3  &s1  &s2  &s3  & s1 &s2  & s3 &s1  &s2  &s3 \\
\hline
\multicolumn{1}{|l|}{M ($M_{\odot}$)}              &1.5  & 2.1 & 1.3 &2.0  &1.2  & 1.7 &1.4  &1.4   &1.8  &1.2  &2.3  &1.7  &1.3 &2.3 &1.9 &1.6 &2.3 &1.2 &1.45&2.0 &1.3 &1.6 &1.4 &1.2\\ 
\multicolumn{1}{|l|}{R (km)}               &13.0 &12.3 & 12.2&14.5 &12.5 & 14.0&10.0 &10.0  &14.5 &9.0  &10.5 &12.0 &6.0 &10.0&8.26&8.0 &10.0&6.0 &12.5&12.0&12.2&10  &7.0 &8.75\\
\multicolumn{1}{|l|}{D (kpc)}              &7.5  & 3.5 & 1.0 &5.8  &6.1  &6.2  &5.5  &5.7   &8    &5.3  &3.5  &7.0  &5.0 &8.0 &6.5 &6.5 &7.0 &5.2 &11.1&8.6 &12.0&10  &7.0 &9.0\\
\multicolumn{1}{|l|}{\textit{i}}      &30   & 13  & 12  &25   &25   &30   &45   &43    &35   &30   &45   &35   &45  &50  &60  &60  &70  &60  &35  &30  &85  &55  &50  &55    \\
\multicolumn{1}{|l|}{$\phi_\mathrm{spot}$ (cycles)} &0.3  & 0.6 & 0.9 &0.3  &0.6  &0.9  &0.3  &0.6   &0.9  &0.8  &0.6  &0.9  &0.3 &0.6 &0.9 &0.3 &0.6 &0.9 &0.3 &0.6 &0.9 &0.0 &0.6 &0.9   \\
\multicolumn{1}{|l|}{$\theta_\mathrm{spot}$ (radian)} &20   & 47  & 42  &25   &20   &22   &20   &20    &20   &25   &25   &20   &65  &55  &60  &55  &70  &60  &40  &60  &40  &55  &50  &55    \\
\multicolumn{1}{|l|}{$\zeta_\mathrm{spot}$ (radian)}   &0.75 & 0.5 & 0.1 &0.34 &0.72 &0.34 &1.0  &1.0   &0.75 &0.9  &0.82 &0.87 &0.75&1.0 &0.75&0.5 &0.51&0.7 &0.75&0.95&0.75&0.44&0.47&0.42 \\
\multicolumn{1}{|l|}{$\log[T_\mathrm{spot}(\mathrm{K})/1\mathrm{K}]$}   &7.2  &7.2  &7.2  &7.3  &7.15 &7.3  &7.19 &7.19  &7.2  &7.18 &7.2  &7.19 &7.36&7.35&7.35&7.38&7.4 &7.34&7.4 &7.5 &7.5 &7.4 &7.45&7.4     \\
\multicolumn{1}{|l|}{$\log[T_\mathrm{star}(\mathrm{K})/1\mathrm{K}]$}      &\none&\none&\none&\none&\none&\none&6.83 &6.83  &6.83 &6.83 &6.83 &6.83 &7.2 &7.2 &7.22&7.2 &7.2 &7.2 &7.21&7.2 &7.2 &7.2 &7.2 &7.2  \\
\multicolumn{1}{|l|}{$N_H$ ($10^{20}\mathrm{cm}^{-2}$)}         & 1   & 3   &6    &7    &2    &2    &2    &5     &3    &2    &8    &8    &1   &9   &5   &6   &6   &6   &1   &5   &5   &1   &1   &1\\

\hline
\caption{Parameters used to produce the for \textit{constant data set}; \none : parameter not used for this case. sX denotes parameter set X  of the corresponding model, X $\in \{1,2,3\}$.
The models indicated correspond to their variability counterparts, based on which synthetic data were generated using the method outlined in Section \ref{subsec:synthetic_data}}.
 \label{tab:params_const}

\end{longtable}
\endgroup

 \clearpage
\twocolumn






\bsp	
\label{lastpage}
\end{document}